\documentclass[12pt]{article}

\usepackage{graphicx}
\usepackage{feynarts}

\makeatletter

\def\paragraph{\@startsection{paragraph}{4}{\z@}{+2.00ex plus
 +1ex minus +.2ex}{1.5ex plus .2ex}{\it\normalsize}}

\def\section{\@startsection {section}{1}{\z@}{+3.0ex plus +1ex minus
  +.2ex}{2.3ex plus .2ex}{\normalsize\bf\boldmath}}
\def\subsection{\@startsection{subsection}{2}{\z@}{+2.5ex plus +1ex
minus +.2ex}{1.5ex plus .2ex}{\normalsize\bf\boldmath}}
\def\subsubsection{\@startsection{subsubsection}{3}{\z@}{+3.25ex plus
 +1ex minus +.2ex}{1.5ex plus .2ex}{\normalsize\it}}

\expandafter\ifx\csname mathrm\endcsname\relax\def\mathrm#1{{\rm #1}}\fi

\renewcommand{\theequation}{\thesection.\arabic{equation}}
\newcounter{saveeqn}

\@addtoreset{equation}{section}

\newcount\@tempcntc
\def\@citex[#1]#2{\if@filesw\immediate\write\@auxout{\string\citation{#2}}\fi
  \@tempcnta\z@\@tempcntb\m@ne\def\@citea{}\@cite{\@for\@citeb:=#2\do
    {\@ifundefined
       {b@\@citeb}{\@citeo\@tempcntb\m@ne\@citea
        \def\@citea{,\penalty\@m\ }{\bf ?}\@warning
       {Citation `\@citeb' on page \thepage \space undefined}}%
    {\setbox\z@\hbox{\global\@tempcntc0\csname
b@\@citeb\endcsname\relax}%
     \ifnum\@tempcntc=\z@ \@citeo\@tempcntb\m@ne
       \@citea\def\@citea{,\penalty\@m}
       \hbox{\csname b@\@citeb\endcsname}%
     \else
      \advance\@tempcntb\@ne
      \ifnum\@tempcntb=\@tempcntc
      \else\advance\@tempcntb\m@ne\@citeo
      \@tempcnta\@tempcntc\@tempcntb\@tempcntc\fi\fi}}\@citeo}{#1}}

\def\@citeo{\ifnum\@tempcnta>\@tempcntb\else\@citea
  \def\@citea{,\penalty\@m}%
  \ifnum\@tempcnta=\@tempcntb\the\@tempcnta\else
   {\advance\@tempcnta\@ne\ifnum\@tempcnta=\@tempcntb \else
\def\@citea{--}\fi
    \advance\@tempcnta\m@ne\the\@tempcnta\@citea\the\@tempcntb}\fi\fi}

\def\nl{\nonumber\\}

\def\asymp#1%
{\mathrel{\raisebox{-.4em}{$\widetilde{\scriptstyle #1}$}}}

\def\Nequal#1%
{\mathrel{\raisebox{-.5em}{$\stackrel{=}{\scriptstyle\rm#1}$}}}
\newcommand{\dsl}[1]{\not \hspace{-0.7mm}#1}
\def\dsl{\mathpalette\make@slash}
\def\make@slash#1#2{\setbox\z@\hbox{$#1#2$}%
  \hbox to 0pt{\hss$#1/$\hss\kern-\wd0}\box0}

\def\beq{\begin{equation}}
\def\eeq{\end{equation}}
\def\beqar{\begin{eqnarray}}
\def\eeqar{\end{eqnarray}}
\def\barr#1{\begin{array}{#1}}
\def\earr{\end{array}}
\def\bfi{\begin{figure}}
\def\efi{\end{figure}}
\def\btab{\begin{table}}
\def\etab{\end{table}}
\def\bce{\begin{center}}
\def\ece{\end{center}}
\def\nn{\nonumber}
\def\disp{\displaystyle}
\def\text{\textstyle}

\def\arraystretch{1.2}

\def\al{\alpha}
\def\be{\beta}
\def\Ga{\Gamma}
\def\ga{\gamma}
\def\de{\delta}
\def\De{\Delta}
\def\eps{\epsilon}

\def\la{\lambda}
\def\om{\omega}

\def\si{\sigma}

\def\refeq#1{\mbox{(\ref{#1})}}
\def\refeqs#1{\mbox{(\ref{#1})}}
\def\refeqf#1{\mbox{(\ref{#1})}}
\def\reffi#1{\mbox{Figure~\ref{#1}}}

\def\refse#1{\mbox{Section~\ref{#1}}}

\def\citere#1{\mbox{Ref.~\cite{#1}}}
\def\citeres#1{\mbox{Refs.~\cite{#1}}}

\newcommand{\TeV}{\unskip\,\mathrm{TeV}}
\newcommand{\GeV}{\unskip\,\mathrm{GeV}}
\newcommand{\MeV}{\unskip\,\mathrm{MeV}}

\newcommand{\fb}{\unskip\,\mathrm{fb}}

\newcommand{\ri}{{\mathrm{i}}}
\newcommand{\rd}{{\mathrm{d}}}
\newcommand{\re}{{\mathrm{e}}}

\newcommand{\rw}{{\mathrm{w}}}

\newcommand{\Oa}{\mathswitch{{\cal{O}}(\alpha)}}

\newcommand{\M}{{\cal{M}}}

\def\mathswitchr#1{\relax\ifmmode{\mathrm{#1}}\else$\mathrm{#1}$\fi}
\newcommand{\Pf}{\mathswitch  f}

\newcommand{\PV}{\mathswitchr V}
\newcommand{\PW}{\mathswitchr W}
\newcommand{\Pw}{\mathswitchr w}
\newcommand{\PZ}{\mathswitchr Z}

\newcommand{\PH}{\mathswitchr H}
\newcommand{\Pe}{\mathswitchr e}

\newcommand{\Pd}{\mathswitchr d}

\newcommand{\Pu}{\mathswitchr u}

\newcommand{\Ps}{\mathswitchr s}

\newcommand{\Pc}{\mathswitchr c}

\newcommand{\Pb}{\mathswitchr b}
\newcommand{\Pbbar}{\mathswitchr{\bar b}}
\newcommand{\Pt}{\mathswitchr t}
\newcommand{\Ptbar}{\mathswitchr{\bar t}}

\newcommand{\Pep}{\mathswitchr {e^+}}
\newcommand{\Pem}{\mathswitchr {e^-}}

\def\mathswitch#1{\relax\ifmmode#1\else$#1$\fi}
\newcommand{\Mf}{\mathswitch {m_\Pf}}

\newcommand{\MV}{\mathswitch {M_\PV}}
\newcommand{\MW}{\mathswitch {M_\PW}}

\newcommand{\MZ}{\mathswitch {M_\PZ}}
\newcommand{\MH}{\mathswitch {M_\PH}}
\newcommand{\Me}{\mathswitch {m_\Pe}}

\newcommand{\Md}{\mathswitch {m_\Pd}}
\newcommand{\Mu}{\mathswitch {m_\Pu}}
\newcommand{\Ms}{\mathswitch {m_\Ps}}
\newcommand{\Mc}{\mathswitch {m_\Pc}}
\newcommand{\Mb}{\mathswitch {m_\Pb}}
\newcommand{\Mt}{\mathswitch {m_\Pt}}

\newcommand{\sw}{\mathswitch {s_\Pw}}
\newcommand{\cw}{\mathswitch {c_\Pw}}

\newcommand{\Qf}{\mathswitch {Q_\Pf}}

\newcommand{\Qt}{\mathswitch {Q_\Pt}}

\newcommand{\GF}{\mathswitch {G_\mu}}

\newcommand{\alphas}{\alpha_{\mathrm{s}}}

\newcommand{\bk}{\mathbf{k}}
\newcommand{\bp}{\mathbf{p}}

\def\ie{i.e.\ }
\def\eg{e.g.\ }

\def\sub{{\mathrm{sub}}}
\def\gsub{g^{(\sub)}}
\def\Gsub{G^{(\sub)}}
\def\cGsub{{\cal G}^{(\sub)}}
\def\cGcoll{{\cal G}^{(\coll)}}

\newcommand{\dsigma}{\ensuremath{\mathrm{d}\sigma}}

\newcommand{\virt}{{\mathrm{virt}}}
\newcommand{\soft}{{\mathrm{soft}}}
\newcommand{\coll}{{\mathrm{coll}}}
\newcommand{\finite}{{\mathrm{finite}}}

\newcommand{\NCt}{N_\Pt^{\mathrm{c}}}

\def\Li{\mathop{\mathrm{Li}_2}\nolimits}

\def\Re{\mathop{\mathrm{Re}}\nolimits}

\def\sgn{\mathop{\mathrm{sgn}}\nolimits}

\newcommand{\eetth}{\Pep\Pem\to\Pt\Ptbar\PH}
\newcommand{\eettha}{\Pep\Pem\to\Pt\Ptbar\PH\ga}

\newcommand{\kH}{k_3}
\newcommand{\KH}{K_3}

\newcommand{\gVtt}{g_{\PV\Pt}^\tau}
\newcommand{\gVtp}{g_{V\Pt}^+}
\newcommand{\gVtm}{g_{V\Pt}^-}
\newcommand{\gVes}{g_{V\Pe}^\si}

\newcommand{\gZtt}{g_{\PZ\Pt}^\tau}

\newcommand{\gZtp}{g_{\PZ\Pt}^+}
\newcommand{\gZtm}{g_{\PZ\Pt}^-}
\newcommand{\gZes}{g_{\PZ\Pe}^\si}

\newcommand{\gHt}{g_{\PH\Pt}}

\newcommand{\gXt}{g_{\chi\Pt}}

\newcommand{\KKtKoK}{\langle k K_2 K_1 k \rangle}

\newcommand{\MKoK}{(k_1\cdot k)}
\newcommand{\MKtK}{(k_2\cdot k)}
\newcommand{\popt} {\langle p_1 p_2  \rangle}
\newcommand{\poeta}{\langle p_1  \eta  \rangle}
\newcommand{\poxi} {\langle p_1  \xi   \rangle}
\newcommand{\ptETA}{\langle p_2  \eta' \rangle}
\newcommand{\ptxi} {\langle p_2 \xi   \rangle}
\newcommand{\poETA}{\langle p_1 \eta' \rangle}
\newcommand{\poXI} {\langle p_1 \xi'  \rangle}
\newcommand{\ptXI} {\langle p_2 \xi'  \rangle}
\newcommand{\pteta}{\langle p_2 \eta  \rangle}
\newcommand{\XIeta} {\langle \xi'  \eta  \rangle}
\newcommand{\xiETA} {\langle \xi   \eta' \rangle}

\newcommand{\Cptxi} {\ptxi^*}
\newcommand{\CptETA}{\ptETA^*}

\newcommand{\CxiETA} {\xiETA^*}
\newcommand{\Cpoxi} {\poxi^*}
\newcommand{\CpoETA}{\poETA^*}

\newcommand{\Kpt}{\langle k p_2 \rangle}
\newcommand{\Kxi} {\langle k \xi \rangle}
\newcommand{\Kpo}{\langle k p_1 \rangle}

\newcommand{\KETA}{\langle k \eta' \rangle}

\newcommand{\CKxi} {\Kxi^*}
\newcommand{\CKpt}{\Kpt^*}

\newcommand{\CKETA}{\KETA^*}
\newcommand{\Cpopt}{\popt^*}                                                   

\newcommand{\ZH}{\mathrm{Z}}
\newcommand{\XH}{\chi}
\newcommand{\VtH}{V\mathrm{t}}
\newcommand{\VtbarH}{V\mathrm{\bar{t}}}

\newcommand{\ZHeA}{\mathrm{Ze}}
\newcommand{\ZHtA}{\mathrm{Zt}}
\newcommand{\XHtA}{\chi\mathrm{t}}
\newcommand{\XHeA}{\chi\mathrm{e}}
\newcommand{\VtHtA}{V\mathrm{tt}}
\newcommand{\VtHeA}{V\mathrm{te}}
\newcommand{\VtbarHtA}{V\mathrm{\bar{t}t}}
\newcommand{\VtbarHeA}{V\mathrm{\bar{t}e}}

\newcommand{\Ln}[2][]{\ln^{#1}\left(#2\right)}
\newcommand{\Dilog}[1]{\Li\left(#1\right)}

\newcommand{\xieta}{\langle\xi  \eta\rangle}
\newcommand{\ETAXI}{\langle\eta'\xi'\rangle}

\def\draftdate{\relax}
\def\mda{\relax}
\def\mua{\relax}
\def\mla{\relax}
\def\draft{
\def\thtystars{******************************}
\def\sixtystars{\thtystars\thtystars}
\typeout{}
\typeout{\sixtystars**}
\typeout{* Draft mode!
         For final version remove \protect\draft\space in source file *}
\typeout{\sixtystars**}
\typeout{}
\def\draftdate{\today}
\def\mua{\marginpar[\boldmath\hfil$\uparrow$]%
                   {\boldmath$\uparrow$\hfil}%
                    \typeout{marginpar: $\uparrow$}\ignorespaces}
\def\mda{\marginpar[\boldmath\hfil$\downarrow$]%
                   {\boldmath$\downarrow$\hfil}%
                    \typeout{marginpar: $\downarrow$}\ignorespaces}
\def\mla{\marginpar[\boldmath\hfil$\rightarrow$]%
                   {\boldmath$\leftarrow $\hfil}%
                    \typeout{marginpar: $\leftrightarrow$}\ignorespaces}
\def\Mua{\marginpar[\boldmath\hfil$\Uparrow$]%
                   {\boldmath$\Uparrow$\hfil}%
                    \typeout{marginpar: $\uparrow$}\ignorespaces}
\def\Mda{\marginpar[\boldmath\hfil$\Downarrow$]%
                   {\boldmath$\Downarrow$\hfil}%
                    \typeout{marginpar: $\downarrow$}\ignorespaces}
\def\Mla{\marginpar[\boldmath\hfil$\Rightarrow$]%
                   {\boldmath$\Leftarrow $\hfil}%
                    \typeout{marginpar: $\leftrightarrow$}\ignorespaces}
\overfullrule 5pt
\oddsidemargin -15mm
\marginparwidth 29mm
}

\def\stars{\strut\leaders\hbox{*}\hfill\strut}
\def\starline{\hfil\strut\hfil\hbox to \textwidth {\stars}\hfil}

\begin{document}
\thispagestyle{empty}
\def\thefootnote{\fnsymbol{footnote}}
\setcounter{footnote}{1}
\null
\draftdate\hfill KA-TP-09-2003\\
\strut\hfill MPP-2003-73 \\
\strut\hfill PSI-PR-03-15\\
\strut\hfill hep-ph/0309274
\vfill
\begin{center}
{\Large \bf\boldmath
Radiative corrections to Higgs-boson production
\\[.5ex]
in association with top-quark pairs at $\Pep\Pem$ colliders
\par} \vskip 2.5em
\vspace{1cm}

{\large
{\sc A.\ Denner$^1$, S.\ Dittmaier$^2$, M. Roth$^3$ and 
M.~M.~Weber$^1$} } \\[1cm]
$^1$ {\it Paul Scherrer Institut, W\"urenlingen und Villigen\\
CH-5232 Villigen PSI, Switzerland} \\[0.5cm]
$^2$ {\it Max-Planck-Institut f\"ur Physik 
(Werner-Heisenberg-Institut) \\
D-80805 M\"unchen, Germany}
\\[0.5cm]
$^3$ {\it Institut f\"ur Theoretische Physik, Universit\"at Karlsruhe \\
D-76128 Karslruhe, Germany}
\par \vskip 1em
\end{center}\par
\vskip 2cm {\bf Abstract:} \par
We have calculated the complete ${\cal O}(\alpha)$ and ${\cal
  O}(\alphas)$ radiative corrections to the Higgs-production process
$\eetth$ in the Standard Model.
This process is particularly interesting for the measurement of the
top-quark Yukawa coupling at a future $\Pep\Pem$ collider.  The
calculation of the ${\cal O}(\alpha)$ corrections is described in some
detail including, in particular, the treatment of the soft and
collinear singularities. The discussion of numerical results focuses
on the total cross section as well as on angular and energy
distributions of the outgoing particles. The electroweak corrections
turn out to be sizable and can reach the order of $\pm 10 \%$.  They
result from cancellations between electromagnetic, fermionic, and weak
bosonic corrections, each of which are of the order of $\pm10 \%$.
\par
\vskip 1cm
\noindent
September 2003
\null
\setcounter{page}{0}
\clearpage
\def\thefootnote{\arabic{footnote}}
\setcounter{footnote}{0}

\section{Introduction}
\label{se:intro}

One of the main tasks at future colliders is the unraveling of the
mechanism of electroweak symmetry breaking.  In the Standard Model
(SM) electroweak symmetry breaking is implemented via the
Higgs--Kibble mechanism, which is responsible for the generation of
particle masses and which leads to the prediction of a physical scalar
particle, the Higgs boson.  The mass of the Higgs boson is expected to
be in the range between the current lower experimental bound of $114.4
\GeV$ \cite{:2001xw} and $1 \TeV$.  Electroweak precision tests yield
a 95\% CL upper limit on the mass of the SM Higgs boson of $211\GeV$
\cite{Grunewald:2003ij}.  A SM Higgs boson in the mass range up to
$1\TeV$ will be discovered at the LHC, provided it exists and has no
exotic properties.  However, for the complete determination of its
profile, including its couplings to fermions and gauge bosons,
experiments in the clean environment of an $\Pep\Pem$ linear collider
\cite{Accomando:1998wt} are indispensable.

The fermion masses in the SM are generated from the Yukawa
interactions via a finite vacuum expectation value
$v=(\sqrt{2}\GF)^{-1/2}\approx246\GeV$ of the Higgs-boson field.  The
strength of the coupling of the physical Higgs boson to fermions is
fixed by $g_{f\bar{f}\PH} = m_f/v$ in lowest order.  The measurement
of these Yukawa couplings is therefore important in order to verify
the mass-generation mechanism of the SM.  By far the largest Yukawa
coupling is the top-quark--Higgs-boson coupling
($g_{\Pt\bar{\Pt}\PH}^2\approx0.5$) owing to the large top-quark mass.
Its measurement is therefore especially interesting.

A direct access to the top-quark Yukawa coupling is provided by the
process $\eetth$ \cite{Gunion:1996vv} if the Higgs-boson mass is not
too large, \ie $\MH\sim 100$--$200\GeV$.  This process proceeds mainly
through Higgs-boson emission off top quarks, while emission from
intermediate Z bosons plays only a minor role, making it suitable for
a determination of $g_{\Pt\bar{\Pt}\PH}$.  If the Higgs boson is
light, \ie $\MH\sim 120\GeV$, a precision of around $5\%$ can be
reached at an $\Pep\Pem$ linear collider operating at $\sqrt{s} =
800\GeV$ with a luminosity of $\int L\,\rd t \sim 1000\fb^{-1}$
\cite{Baer:1999ge}.  Combining the $\Pt\bar{\Pt}\PH$ channel with
information from other Higgs-production and decay processes an even
better accuracy can be obtained in a combined fit
\cite{Battaglia:2000jb}.  Furthermore, by investigating the process
$\eetth$ bounds on non-standard physics in the top-quark Yukawa
coupling can be derived \cite{Gunion:1996vv,Han:1999xd}.

Assuming an experimental precision of a few per cent, a thorough
understanding of the background to the $\Pt\bar{\Pt}\PH$ final state
is necessary \cite{Moretti:1999kx}, and precise theoretical
predictions for the signal process $\eetth$ at the per-cent level are
needed. This requires the inclusion of radiative corrections, a rather
complicated task for a process with three massive unstable particles
in the final state.  As a first step, the process $\eetth$ can be
treated in the approximation of stable top quarks and Higgs bosons.
The corresponding leading-order total cross section is already known
for a long time \cite{Gaemers:1978jr}.  The ${\cal O}(\alphas)$
corrections to the total cross section within the SM have been
calculated first in the ``effective Higgs-boson approximation'' valid
only for small Higgs-boson masses and very high energies
\cite{Dawson:1997im}.  A calculation of all diagrams for the dominant
photon-exchange channel has been performed in \citere{Dawson:1998ej},
while the full set of diagrams has been evaluated in
\citere{Dittmaier:1998dz}. Within the Minimal Supersymmetric Standard
Model (MSSM) the ${\cal O}(\alphas)$ corrections to the production
cross sections of neutral Higgs bosons in association with heavy quark
pairs ($\Pt\Ptbar$, $\Pb\Pbbar$) have been calculated in
\citere{Dawson:1998qq} for the photon-exchange channel. In
\citere{Dittmaier:2000tc} all QCD diagrams have been taken into
account, while the SUSY-QCD corrections have been worked out in
\citere{Zhu:2002iy}.  Furthermore, the ${\cal O}(\alphas)$ corrections
to the Higgs-boson energy distribution both in the SM and the MSSM
have been discussed in \citere{Dittmaier:mg}.

Recently, considerable progress has been achieved in the calculation
of the electroweak corrections to $\eetth$.  Results for the
electroweak $\Oa$ corrections in the SM have been presented in
\citeres{You:2003zq,Belanger:2003nm}.  A calculation of some top-mass
enhanced electroweak corrections in the SM, the MSSM, and the
two-Higgs doublet model has been performed in \citere{Wu:2003sp}.

We have accomplished a completely independent calculation of the $\Oa$
electroweak corrections to the process $\eetth$ in the SM as well as
of the ${\cal O}(\alphas)$ QCD corrections.  First results for total
cross sections have already been presented in \citere{Denner:2003ri}
including a comparison of our results with those of
\citeres{You:2003zq,Belanger:2003nm}. While we find good agreement
with \citere{Belanger:2003nm}, our results differ from those of
\citere{You:2003zq} at high centre-of-mass (CM) energies and close to
threshold.  Moreover, we reproduced the results of
\citere{Dittmaier:1998dz} for the QCD corrections to this process.

In this paper we present details of our calculation.  Single
hard-photon radiation is included using the complete matrix element
and combined with the virtual corrections applying both
phase-space-slicing and subtraction methods.  Higher-order corrections
from initial-state photon radiation (ISR) are taken into account in
the leading-logarithmic approximation.  Moreover, we supplement our
previous work with results for distributions in the energies and
production angles of the Higgs boson and the top quark.  Results on
electroweak corrections to differential cross sections have not yet
been presented in the literature before.

The paper is organized as follows: in \refse{se:calcrcs} the
calculation of the virtual and real corrections is described.  Section
\ref{se:numres} contains a discussion of numerical results. The paper
is summarized in \refse{se:sum}.  Further useful information on the
calculation of the matrix elements is collected in the appendices.

\section{Calculation of radiative corrections}
\label{se:calcrcs}

\subsection{Conventions and lowest-order cross section}
\label{se:convs}

We consider the process
\beq
\Pem(p_1,\sigma_1) + \Pep(p_2,\sigma_2) \;\longrightarrow\;
\Pt(k_1, \tau_1) + \Ptbar(k_2, \tau_2) + \PH(\kH),
\label{eq:eennh}
\eeq 
where the momenta $p_a$, $k_i$ of the particles and the helicities of
the electrons, $\sigma_a$, and top quarks, $\tau_i$, are given in
parentheses.  The helicities take the values $\sigma_a,\tau_i=\pm1/2$,
but we often use only the sign to indicate the helicity.  The electron
mass is neglected whenever possible, \ie it is kept finite only in the
mass-singular logarithms related to ISR. This implies that the
lowest-order and one-loop amplitudes vanish unless $\si_1=-\si_2$.
Therefore, we define $\si=\si_1=-\si_2$.  The particle momenta obey
the mass-shell conditions $p_1^2=p_2^2=0$, $k_1^2=k_2^2=\Mt^2$, and
$\kH^2=\MH^2$.  For later use, the following set of kinematical
invariants is defined:
\beqar
s &=& (p_1+p_2)^2, \nn\\
s_{ij} &=& (k_i+k_j)^2, \qquad i,j=1,2,3, \nn\\
t_{ai} &=& (p_a-k_i)^2, \qquad a=1,2, \quad i=1,2,3.  
\eeqar

Using the conventions of \citere{Denner:1993kt} we generically denote
the couplings of the neutral bosons to the fermions in the following
by
\beqar
g_{\gamma f}^\pm &=& -\Qf, \qquad
g_{\PZ f}^+ = -\frac{\sw}{\cw}\Qf, \qquad
g_{\PZ f}^- = -\frac{\sw}{\cw}\Qf + \frac{I^3_{\rw,f}}{\cw\sw},
\nn\\
g_{\PH f}^\pm &=& g_{\PH f} = -\frac{1}{2\sw}\frac{\Mf}{\MW}, \qquad
g_{\chi f}^\pm = \pm g_{\chi f} = \pm\frac{\ri I^3_{\rw,f}}{\sw}\frac{\Mf}{\MW},
\nl
g_{\PZ\PZ\PH}&=&\frac{\MZ}{\sw\cw},\qquad g_{\PZ\chi\PH}=\frac{-\ri}{2\sw\cw},
\label{eq:couplings}
\eeqar
where $\Qf$ is the relative charge of the fermion $f$, and
$I^3_{\rw,f}=\pm{1}/{2}$ the third component of the weak isospin of
the left-handed part of the fermion field $f$.  The sine and cosine of
the weak mixing angle are fixed by
\beq\label{eq:defcw}
\cw^2 = 1-\sw^2 = \frac{\MW^2}{\MZ^2}.
\eeq

In order to express the amplitude for the bremsstrahlung process
$\eettha$ in a compact way (see \refse{se:bremsme}), we use the
Weyl--van der Waerden spinor technique as formulated in
\citere{Dittmaier:1999nn}.  The helicity states for the massless
electrons and massive top quarks are constructed as follows. Those for
the massless incoming electron with momentum
\beq
p_1^\mu=p_1^0(1,\cos\phi_{p_1}\sin\theta_{p_1},
\sin\phi_{p_1}\sin\theta_{p_1},\cos\theta_{p_1})
\eeq
can be constructed from the Weyl spinor
\beq\label{eq:electronspinor}
p_{1,A} =
\sqrt{2p_1^0}\pmatrix{\re^{-\ri\phi_{p_1}}\cos\frac{\theta_{p_1}}{2}
\cr\sin\frac{\theta_{p_1}}{2}},
\eeq
and the corresponding momentum matrix reads
\beq\label{eq:electronmomentummatrix}
P_{1,\dot AB} = p_{1,\dot A} p_{1,B}.  
\eeq
The spinor $p_{2,A}$ and the momentum matrix $P_{2,\dot AB}$ of the
incoming positron are defined analogously.

The spinors $\rho_{i,A}$ corresponding to the top-quark momentum
\beq
k_1^\mu=(k^0_1,|\bk_1|\cos\phi_{k_1}\sin\theta_{k_1},
|\bk_1|\sin\phi_{k_1}\sin\theta_{k_1},|\bk_1|\cos\theta_{k_1})
\eeq
are defined as 
\beqar\label{massivespinors}
\rho_{1,A} &=&
\sqrt{k^0_1+|\bk_1|}\pmatrix{\re^{-\ri\phi_{k_1}}
\cos\frac{\theta_{k_1}}{2}\cr\sin\frac{\theta_{k_1}}{2}}, \qquad
\rho_{2,A} =
\sqrt{k^0_1-|\bk_1|}\pmatrix{\sin\frac{\theta_{k_1}}{2}\cr 
-\re^{+\ri\phi_{k_1}}\cos\frac{\theta_{k_1}}{2}}.
\label{eq:momwvdw}
\eeqar
They obey the relation
\beq
\langle\rho_2\rho_1\rangle = \Mt,
\eeq
where the spinor product is defined by
\beq
\langle \phi\psi\rangle = \phi_A\eps^{AB}\phi_B=\phi_1\psi_2- \phi_2\psi_1.
\eeq
The spinors $\rho'_{i,A}$ corresponding to $k_2$ are constructed in
the same way.  The momentum matrices $K_1$ and $K_2$ corresponding the
outgoing momenta $k_1$ and $k_2$ are decomposed into the spinors
$\rho_{i,A}$ and $\rho'_{i,A}$, respectively,
\beq
K_{1,\dot AB} = \sum_{i=1,2} \rho_{i,\dot A} \rho_{i,B}, \qquad
K_{2,\dot AB} = \sum_{i=1,2} \rho'_{i,\dot A} \rho'_{i,B}.
\eeq
The Dirac spinors of the outgoing (anti-)top quarks are generically
given by
\beq
\bar u_\Pt(k_1) =
(\eta^A,\xi_{\dot A}),
\qquad
v_{\bar\Pt}(k_2) =
\pmatrix{ \vphantom{\xi'_{\dot A}}\xi'_A \cr \eta^{\prime\dot A} }
\eeq
with the actual insertions
\beq\label{eq:etaxi}
(\xi,\eta) = \Biggl\{
\barr{ll} (\rho_1,-\rho_2) & \quad \mbox{for} \; \tau_1=+, \\
          (\rho_2, \rho_1) & \quad \mbox{for} \; \tau_1=-, \earr
\qquad
(\xi',\eta') = \Biggl\{
\barr{ll} ( \rho'_1,\rho'_2) & \quad \mbox{for} \; \tau_2=-, \\
          (-\rho'_2,\rho'_1) & \quad \mbox{for} \; \tau_2=+, \earr
\eeq
\ie we have $\xieta=\ETAXI=\Mt$. 

Although the Higgs boson has spin 0, it is convenient to express its
momentum in terms of Weyl spinors $\phi_{i,A}$.  These are obtained
from the momentum $\kH$ analogously to \refeq{massivespinors}.  Then
the momentum matrix $\KH$ of the Higgs boson can be decomposed as
\beq
K_{3,\dot AB} = \sum_{i=1,2} \phi_{i,\dot A} \phi_{i,B}.
\eeq

It is instructive to give the  lowest-order amplitude for $\eetth$ in
this formalism. 
\begin{figure}
\centerline{\footnotesize  \input{paper-tree}}
\vspace{-.5em}
\caption{Tree-level diagrams for $\eetth$}
\label{fi:treediags}
\end{figure}
It receives contributions of the four diagrams shown in
\reffi{fi:treediags}:
\beq\label{eq:MEttH}
\M_0^{\si \tau_1 \tau_2} = 
\M_0^{\ZH,\si \tau_1 \tau_2} 
+\M_0^{\XH,\si \tau_1 \tau_2} 
+\sum_{V=\ga,\PZ}\left(\M_0^{\VtH,\si \tau_1 \tau_2} 
+\M_0^{\VtbarH,\si \tau_1 \tau_2}\right),
\eeq
where the upper indices $\PZ,\chi,\Pt,\Ptbar$ indicate the particle
from which the Higgs boson is emitted.  In the 't~Hooft--Feynman
gauge, the contributions corresponding to these four diagrams read
\beqar\label{eq:MEttHi}
\M_0^{\ZH,\si \tau_1 \tau_2} &=& 
{2\,e^3\, \gZes\, g_{\PZ\PZ\PH}} \,
P_\PZ(p_1+p_2) \, P_\PZ(k_1+k_2) \,
A_{\si\tau_1\tau_2}^{\ZH}(p_1,p_2,k_1,k_2),
\nl[.5em]
\M_0^{\XH,\si \tau_1 \tau_2} &=& 
{-2e^3 \gZes\, g_{\PZ\chi\PH}\,\gXt} \,
P_\PZ(p_1+p_2) \,
P_\PZ(k_1+k_2) \,
A_{\si\tau_1\tau_2}^{\XH}(p_1,p_2,k_1,k_2),
\nl[.5em]
\M_0^{\VtH,\si \tau_1 \tau_2} &=& 
-2\,e^3\,\gVes\,\gHt \,
P_\PV(p_1+p_2) \,
P_\Pt(k_1+\kH) \,
A_{\si\tau_1\tau_2}^{\VtH}(p_1,p_2,k_1,k_2),
\nl[.5em]
\M_0^{\VtbarH,\si \tau_1 \tau_2} &=& 
-2\,e^3\,\gVes\,\gHt \,
P_\PV(p_1+p_2) \,
P_\Pt(k_2+\kH) \,
A_{\si\tau_1\tau_2}^{\VtbarH}(p_1,p_2,k_1,k_2)
\eeqar
with the auxiliary functions
\beqar\label{eq:MEttHaux}
A_{+,\tau_1,\tau_2}^{\ZH}(p_1,p_2,k_1,k_2) &=&
\gZtp\Cptxi\poXI + \gZtm\CptETA\poeta, 
\nl[.5em]
A_{+,\tau_1,\tau_2}^{\XH}(p_1,p_2,k_1,k_2) &=& 
\Bigl(\XIeta-\CxiETA \Bigr)
\langle p_2\KH p_1\rangle,
\nl[.5em]
A_{+,\tau_1,\tau_2}^{\VtH}(p_1,p_2,k_1,k_2) &=&
\gVtm\CptETA\Bigl(2\Mt\poeta+\langle \xi\KH p_1\rangle\Bigr)
\nl&&{}
+\gVtp\poXI\Bigl(2\Mt\Cptxi-\langle p_2\KH\eta\rangle\Bigr),
\nl[.5em]
A_{+,\tau_1,\tau_2}^{\VtbarH}(p_1,p_2,k_1,k_2) &=&
\gVtp\Cptxi\Bigl(2\Mt\poXI+\langle \eta'\KH p_1\rangle\Bigr)
\nl&&{}
+\gVtm\poeta\Bigl(2\Mt\CptETA-\langle p_2\KH\xi'\rangle\Bigr),
\nn\\[.5em]
A_{-\si,\tau_1,\tau_2}^{\ldots}(p_1,p_2,k_1,k_2) &=&
A_{\si\tau_1\tau_2}^{\ldots}(p_2,p_1,k_1,k_2).
\eeqar
Here we used the shorthand notation
\beq
\label{eq:auxkh}
\langle \psi' \KH \psi\rangle =
\psi'_{\dot A} \KH^{\dot AB} \psi_{B} = \sum_{i=1,2}
\langle \psi' \phi_i \rangle^* \langle \psi \phi_i \rangle
\eeq
and abbreviated the propagators as
\beq
P_{\Pt}(p)=\frac{1}{p^2-\Mt^2},\qquad
P_{\PZ}(p)=\frac{1}{p^2-\MZ^2}, \qquad 
P_{\gamma}(p)=\frac{1}{p^2}.
\eeq
Note that there is no need to introduce a finite width since none of
the internal lines can become resonant for the physical top-quark and
Higgs-boson masses.  Finally, the lowest-order cross section reads
\beqar\label{eq:sigma0}
\sigma_0 &=& \frac{1}{2s} \,
\int\rd\Phi_3 \sum_{\si=\pm\frac12}
\frac{1}{4}(1+2P_-\si)(1-2P_+\si) 
|\M_0^{\si}|^2,
\eeqar
where
\beqar
|\M_0^{\si}|^2&=&\NCt\sum_{\tau_1=\pm\frac12} \sum_{\tau_2=\pm\frac12}
|\M_0^{\si\tau_1\tau_2}|^2
\eeqar
with the colour factor $\NCt=3$ of the top quark, $P_\pm$ are the
degrees of polarization of the $\Pe^\pm$ beams, and the integral over
the three-particle phase space is defined by
\beq
\int \rd\Phi_3 =
\left( \prod_{i=1}^3 \int\frac{\rd^3 {\bf k}_i}{(2\pi)^3 2k_i^0} \right)\,
(2\pi)^4 \delta\Biggl(p_1+p_2-\sum_{j=1}^3 k_j\Biggr).
\label{eq:dG3}
\eeq

\subsection{Virtual corrections}
\label{se:vrcs}

\subsubsection{Survey of one-loop diagrams}

The virtual corrections receive contributions from self-energy,
vertex, box, and pentagon diagrams.  The structural diagrams
containing the generic contributions of vertex functions are
summarized in \reffi{fi:gendiagrams}.  The full set of pentagon
diagrams is shown in \reffi{fi:eetth}.
\begin{figure}
\centerline{\footnotesize  \input{paper-vertex}}
\vspace{-.5em}
\caption{Contributions of different vertex functions to $\eetth$}
\label{fi:gendiagrams}
\end{figure}%
\begin{figure}
\centerline{\footnotesize  \input{paper-eetth}}
\vspace{-.5em}
\caption{Pentagon diagrams for $\eetth$}
\label{fi:eetth}
\end{figure}

\begin{figure}
\centerline{\footnotesize  \input{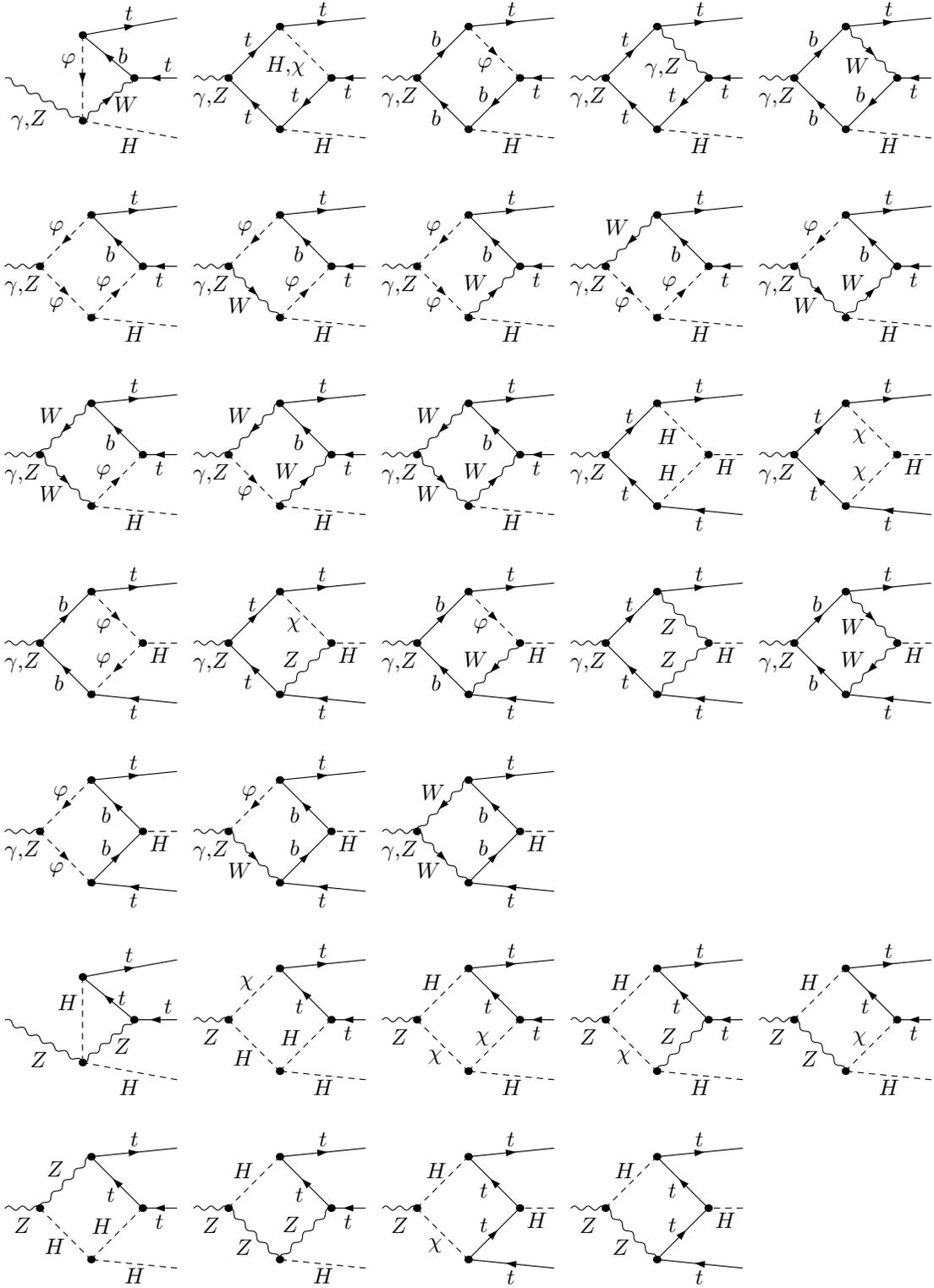}}
\vspace{-.5em}
\caption{$\gamma\Pt\Ptbar\PH$ and $\PZ\Pt\Ptbar\PH$ box diagrams 
(Diagrams with reversed charge flow are not shown.)}
\label{fi:ztth}
\end{figure}

\begin{figure}
\centerline{\footnotesize  \input{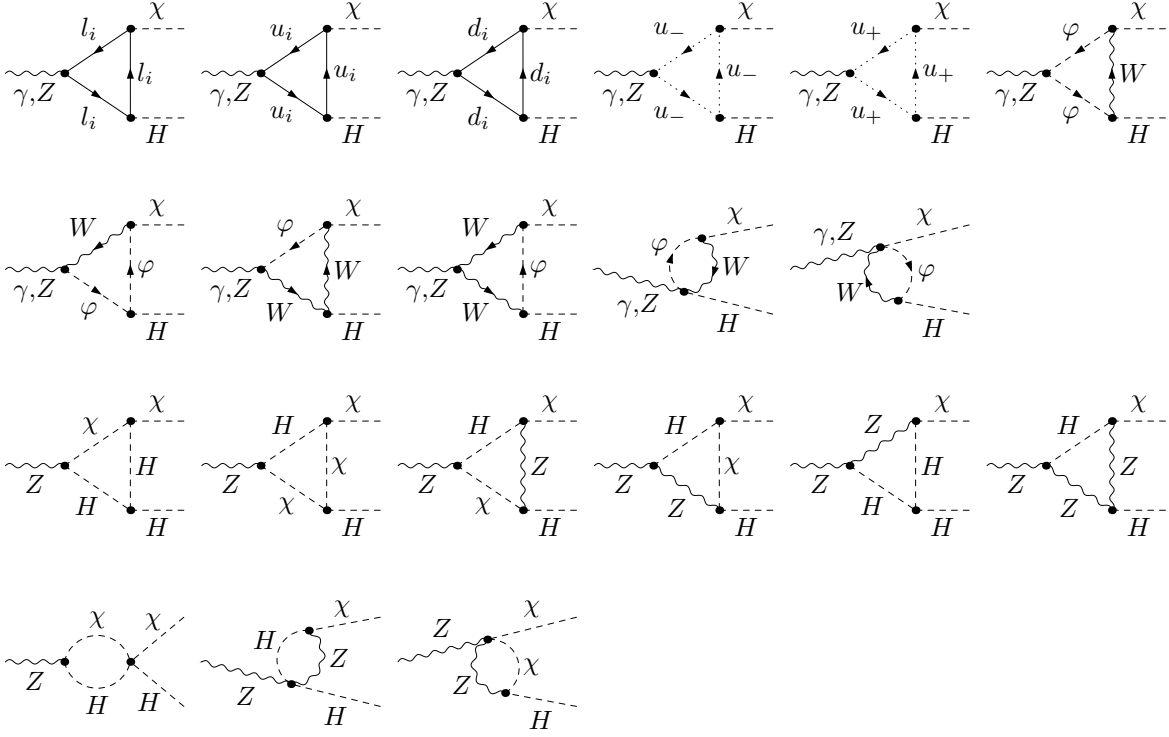}}
\vspace{-.5em}
\caption{Diagrams for $\gamma\PH\chi$ and $\PZ\PH\chi$ vertex functions
  (Diagrams with reversed charge flow in the loop are not shown.)}
\label{fi:zhg0}
\end{figure}

The box diagrams for the $\Pep\Pem\Pt\Ptbar$ vertex function are shown
in \reffi{fi:eett}, the $\Pep\Pem\PH\PH$ box diagrams in
\reffi{fi:eehh}, and the $\Pep\Pem\PH\chi$ box diagrams in
\reffi{fi:eehg0}.  \reffi{fi:ztth} lists the diagrams for the
$\gamma\Pt\Ptbar\PH$ and $\PZ\Pt\Ptbar\PH$ vertex functions.  In
\reffi{fi:ztth}, as well as in \reffi{fi:zhg0}, diagrams that can be
obtained by reversing the charge flow of all charged particles are not
shown.
\begin{figure}
\centerline{\footnotesize  \input{paper-eett}}
\vspace{-.5em}
\caption{$\Pep\Pem\Pt\Ptbar$ box diagrams}
\label{fi:eett}
\end{figure}%
\begin{figure}
\centerline{\footnotesize  \input{paper-eehh}}
\vspace{-.5em}
\caption{$\Pep\Pem\PH\PH$ box diagrams}
\label{fi:eehh}
\end{figure}%
\begin{figure}
\centerline{\footnotesize  \input{paper-eehg0}}
\vspace{-.5em}
\caption{$\Pep\Pem\PH\chi$ box diagrams}
\label{fi:eehg0}
\end{figure}%
The $\Pep\Pem\PZ\PH$ box diagrams can be found in
\citere{Denner:2003yg}, and the $\Pep\Pem\gamma\PH$ diagrams can be
obtained from a subset of those.

\begin{figure}
\centerline{\footnotesize  \input{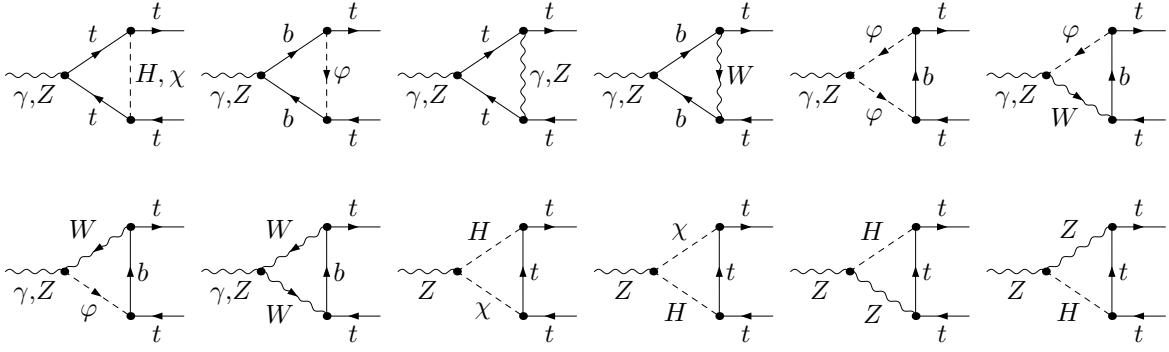}}
\vspace{-.5em}
\caption{Diagrams for $\gamma\Pt\Ptbar$ and $\PZ\Pt\Ptbar$ vertex functions}
\label{fi:ztt}
\end{figure}

\begin{figure}
\centerline{\footnotesize  \input{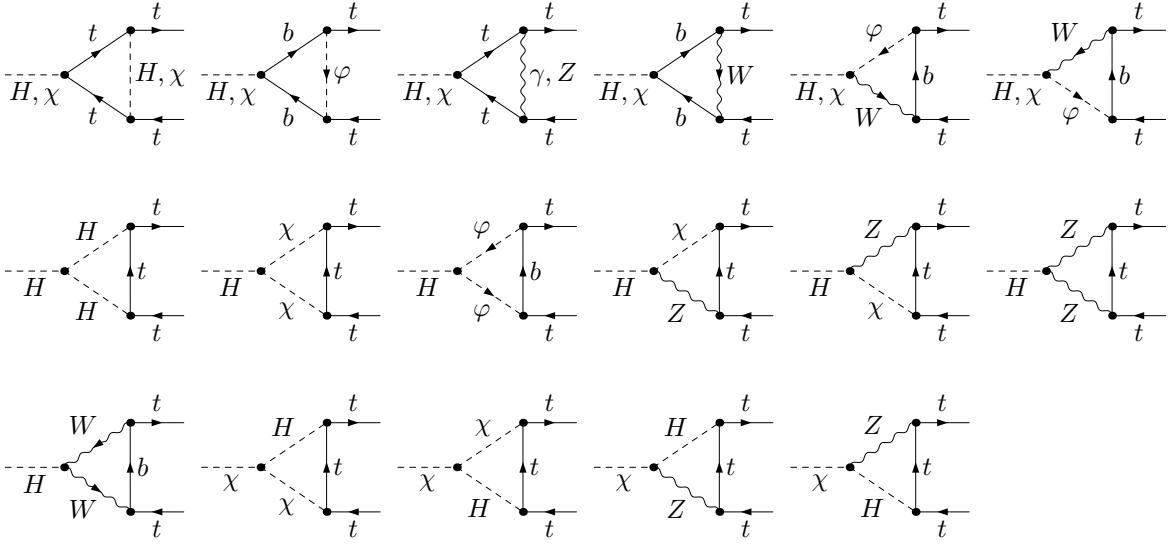}}
\vspace{-.5em}
\caption{Diagrams for $\PH\Pt\Ptbar$ and $\chi\Pt\Ptbar$ vertex functions}
\label{fi:htt}
\end{figure}

\begin{figure}
\centerline{\footnotesize  \input{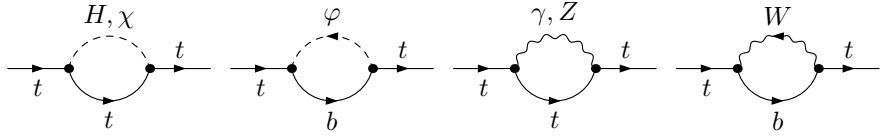}}
\vspace{-.5em}
\caption{Diagrams contributing to the top-quark self-energy}
\label{fi:tt}
\end{figure}

The diagrams contributing to the $\gamma\Pt\Ptbar$ and $\PZ\Pt\Ptbar$
vertex functions are depicted in \reffi{fi:ztt}, the $\PH\Pt\Ptbar$
and $\chi\Pt\Ptbar$ diagrams in \reffi{fi:htt}, and \reffi{fi:zhg0}
shows the $\gamma\PH\chi$ and $\PZ\PH\chi$ diagrams.
The diagrams for the $\gamma\PZ\PH$, $\PZ\PZ\PH$, and $\Pep\Pem\PH$
vertex functions are shown in \citere{Denner:2003yg}. Those for the
$\gamma\gamma\PH$ vertex function can be obtained from a subset of the
$\gamma\PZ\PH$ vertex diagrams.  The $\Pep\Pem\gamma$ and
$\Pep\Pem\PZ$ diagrams and most of the self-energy diagrams can be
found in \citere{Hollik:1988ii}.  The diagrams for the top-quark
self-energy are listed in \reffi{fi:tt}.
The $\PZ\PH$-mixing energy and the $\ga\PH\PH$ and $\PZ\PH\PH$ vertex
functions vanish owing to CP symmetry.

All pentagon and box diagrams are ultraviolet (UV) finite, as well as
the $\gamma\gamma\PH$ vertex function.  Since we neglect the electron
mass whenever possible, also the $\Pep\Pem\PH$ vertex function is UV
finite.  The remaining vertex functions and self-energies are UV
divergent, and the corresponding counterterm diagrams have to be
included.

\subsubsection{Calculational framework}

The actual calculation of the one-loop diagrams closely follows the
strategy of \citere{Denner:2003yg}, where the ${\cal O}(\alpha)$
corrections to $\Pep\Pem\to\nu\bar\nu\PH$ have been calculated, i.e.\ 
it has been carried out in the 't~Hooft--Feynman gauge using standard
techniques.  The Feynman graphs have been generated with {\sl
  FeynArts} \cite{Kublbeck:1990xc} and are evaluated in two completely
independent ways, leading to two independent computer codes.  The
results of the two codes are in good numerical agreement (i.e.\ within
about 12 digits for non-exceptional phase-space points).  Apart from
the 5-point integrals, the tensor coefficients of the one-loop
integrals are algebraically reduced to scalar integrals with the
Passarino--Veltman algorithm \cite{Passarino:1979jh} at the numerical
level. For the evaluation of the tensor 5-point functions the direct
reduction to 4-point integrals of \citere{Denner:2002ii} has been
used. This method avoids leading inverse Gram determinants which cause
numerical instabilities at the phase-space boundaries. The scalar
integrals are evaluated using the methods and results of
\citeres{Denner:1993kt,'tHooft:1979xw}, where UV divergences are
regulated dimensionally and infrared (IR) divergences with an
infinitesimal photon mass $m_\gamma$.  The soft and collinear
singularities of some scalar 5-point integrals have been checked
against the general result of \citere{Dittmaier:2003bc}.  The
renormalization is carried out in the on-shell renormalization scheme,
as \eg described in \citere{Denner:1993kt}.

In the first calculation, the Feynman graphs are generated with {\sl
  FeynArts} version~1.0 \cite{Kublbeck:1990xc}. With the help of
{\sl Mathematica} routines the amplitudes are expressed in terms of
standard matrix elements (SME) and coefficients of tensor integrals,
as described in the appendices in more detail.  The output is
processed into a {\sl Fortran} program for the numerical evaluation.
The second calculation has been made using {\sl FeynArts} version~3
\cite{Hahn:2000kx} for the generation and {\sl FormCalc}
\cite{Hahn:1998yk} for the evaluation of the amplitudes. The
analytical results of {\sl FormCalc} in terms of SME and their
coefficients were translated into {\sl C{+}{+}} code for numerical
evaluation.

As a check of gauge invariance the calculation of the virtual
corrections has been repeated using the background-field method
\cite{Denner:1994xt}, where the individual contributions from
self-energy, vertex, box, and pentagon corrections differ from their
counterparts in the conventional formalism.  The total one-loop
corrections of the conventional and of the background-field approach
were found to be in perfect numerical agreement.

Finally, the contribution of the virtual corrections to the cross
section is given by
\beq
\de\sigma_{\virt} = \frac{1}{2s} \,
\int\rd\Phi_3 \sum_{\si=\pm\frac12}
\frac{1}{4}(1+2P_-\si)(1-2P_+\si) \,\NCt
\sum_{\tau_1=\pm\frac12} \sum_{\tau_2=\pm\frac12}
2\Re\left\{\M_1^{\si\tau_1\tau_2}\left(\M_0^{\si\tau_1\tau_2}\right)^*\right\},
\label{eq:sigmaV}
\eeq
where $\M_1^{\si\tau_1\tau_2}$ denotes the one-loop helicity
amplitudes.

\subsection{Real photonic corrections}
\label{se:rrcs}

\subsubsection{Matrix-element calculation}
\label{se:bremsme}

The real photonic corrections are induced by the process
\beq
\Pem(p_1,\sigma_1) + \Pep(p_2,\sigma_2) \;\longrightarrow\;
\Pt(k_1,\tau_1) + \bar\Pt(k_2,\tau_2) + \PH(\kH) + \gamma(k,\lambda), 
\label{eq:eettha}
\eeq
where $k$ and $\lambda$ denote the photon momentum and helicity,
respectively. The Feynman diagrams of this process are shown in
\reffi{fi:eettha}.
\begin{figure}
\centerline{\footnotesize  \input{paper-real}}
\vspace{-.5em}
\caption{Feynman diagrams for $\Pep\Pem\to\Pt\bar\Pt\PH\ga$}
\label{fi:eettha}
\end{figure}
We have evaluated the corresponding helicity matrix elements
$\M^{\si\tau_1\tau_2\la}_\ga$ using the Weyl--van der Waerden spinor
technique as formulated in \citere{Dittmaier:1999nn}.  The amplitudes
for the helicity channels with $\si_1=\si_2$ vanish for massless
electrons, and we define $\si=\si_1=-\si_2$ as above.  The treatment
of IR (soft and collinear) singularities, where the finite electron
mass is needed, is described below.

The amplitude for $\eettha$ can be separated into 12 contributions
that are invariant with respect to gauge transformations of the
external photon,
\beqar\label{eq:MEttHg}
\M^{\si \tau_1 \tau_2 \lambda}_\ga &=& \M^{\ZHeA,\si \tau_1 \tau_2 \la}_{\ga} 
+ \M^{\ZHtA,\si \tau_1 \tau_2 \la}_{\ga} 
+ \M^{\XHeA,\si \tau_1 \tau_2 \la}_{\ga} 
+ \M^{\XHtA,\si \tau_1 \tau_2 \la}_{\ga} 
\nl &&{}
+ \sum_{V=\ga,\PZ}\left(\M^{\VtHeA,\si \tau_1 \tau_2 \la}_{\ga} 
+ \M^{\VtHtA,\si \tau_1 \tau_2 \la}_{\ga} 
+ \M^{\VtbarHeA,\si \tau_1 \tau_2 \la}_{\ga} 
+ \M^{\VtbarHtA,\si \tau_1 \tau_2 \la}_{\ga} \right).
\eeqar
The last upper particle index in $\M^{\ZHeA}_\ga$, etc., indicates to
which charged fermion line \mbox{($\Pe$ or $\Pt$)} the photon is
attached.  The diagrams in the first line of \reffi{fi:eettha} belong
to $\M^{\ZHeA}_\ga$ and $\M^{\ZHtA}_\ga$, those in the second line to
$\M^{\XHeA}_\ga$ and $\M^{\XHtA}_\ga$, those in the third line to
$\M^{\VtHeA}_\ga$ and $\M^{\VtHtA}_\ga$, and those in the last line to
$\M^{\VtbarHeA}_\ga$ and $\M^{\VtbarHtA}_\ga$.

In the `t~Hooft--Feynman gauge, the individual contributions read
\beqar\label{eq:MEttHgi}
\M^{\ZHeA,\si \tau_1 \tau_2 \la}_{\ga} &=& 
{2\sqrt{2}\,e^4\, \gZes\, g_{\PZ\PZ\PH}} \,
P_\PZ(p_1+p_2-k) \,
P_\PZ(k_1+k_2) \,
\nl*&& {} \times 
A_{\si\tau_1\tau_2\la}^{\ZHeA}(p_1,p_2,k_1,k_2,k),
\\[.5em]
\label{eq:ZHtA}
\M^{\ZHtA,\si \tau_1 \tau_2 \la}_{\ga} &=& 
{-\sqrt{2}\,e^4\, \Qt\,\gZes\, g_{\PZ\PZ\PH}} \,
P_\PZ(p_1+p_2) \,
P_\PZ(k_1+k_2+k) \,
\nl*&& {} \times 
A_{\si\tau_1\tau_2\la}^{\ZHtA}(p_1,p_2,k_1,k_2,k),
\\[.5em]
\M^{\XHeA,\si \tau_1 \tau_2 \la}_{\ga} &=& 
{2\sqrt{2}\,e^4 \gZes\, g_{\PZ\chi\PH}\,\gXt} \,
P_\PZ(p_1+p_2-k) \,
P_\PZ(k_1+k_2) \,
\nl* && {} \times 
A_{\si\tau_1\tau_2\la}^{\XHeA}(p_1,p_2,k_1,k_2,k),
\\[.5em]
\label{eq:XHtA}
\M^{\XHtA,\si \tau_1 \tau_2 \la}_{\ga} &=& 
{-2\sqrt{2}\,e^4\, Q_\Pt \gZes \,g_{\PZ\chi\PH}\,\gXt} \,
P_\PZ(p_1+p_2) \,
P_\PZ(k_1+k_2+k) \,
\nl* && {} \times 
A_{\si\tau_1\tau_2\la}^{\XHtA}(p_1,p_2,k_1,k_2,k),
\\[.5em]
\M^{\VtHeA,\si \tau_1 \tau_2 \la}_{\ga} &=& 
2\sqrt{2}\,e^4\,\gVes\,\gHt \,
P_\PV(p_1+p_2-k) \,
P_\Pt(k_1+\kH) \,
\nl* && {} \times 
A_{\si\tau_1\tau_2\la}^{\VtHeA}(p_1,p_2,k_1,k_2,k),
\\[.5em]
\label{eq:VtHtA}
\M^{\VtHtA,\si \tau_1 \tau_2 \la}_{\ga} &=& 
-2\sqrt{2}\,e^4\,\Qt\,\gVes\,\gHt \,
P_\PV(p_1+p_2) \,
A_{\si\tau_1\tau_2\la}^{\VtHtA}(p_1,p_2,k_1,k_2,k),
\\[.5em]
\M^{\VtbarHeA,\si \tau_1 \tau_2 \la}_{\ga} &=& 
2\sqrt{2}\,e^4\,\gVes\,\gHt \,
P_\PV(p_1+p_2-k) \,
P_\Pt(k_2+\kH) \,
\nl && {} \times 
A_{\si\tau_1\tau_2\la}^{\VtbarHeA}(p_1,p_2,k_1,k_2,k),
\\[.5em]
\label{eq:VtbarHtA}
\M^{\VtbarHtA,\si \tau_1 \tau_2 \la}_{\ga} &=& 
-2\sqrt{2}\,e^4\,\Qt\,\gVes\,\gHt \,
P_\PV(p_1+p_2) \,
A_{\si\tau_1\tau_2\la}^{\VtbarHtA}(p_1,p_2,k_1,k_2,k)
\eeqar
with the auxiliary functions
\beqar\label{eq:MEttHgaux}
\lefteqn{A_{+,\tau_1,\tau_2,+}^{\ZHeA}(p_1,p_2,k_1,k_2,k) =
 -\gZtp\poXI\frac{\langle \xi(P_2-K)p_1\rangle}{\Kpo\Kpt}
 -\gZtm\poeta\frac{\langle \eta'(P_2-K)p_1\rangle}{\Kpo\Kpt},}\quad
\nl[.5em]
\lefteqn{A_{+,\tau_1,\tau_2,+}^{\ZHtA}(p_1,p_2,k_1,k_2,k) =
 \Bigl(\gZtp\Cptxi\poXI + \gZtm\CptETA\poeta\Bigr)
\frac{\KKtKoK}{2\MKoK\MKtK}}\quad\nl*
&& {} + 
 \CKpt\left(\frac{\gZtp\CKxi}{\MKoK} \poXI
- \frac{\gZtm\CKETA}{\MKtK} \poeta\right),
\nl[.5em]
\lefteqn{A_{+,\tau_1,\tau_2,+}^{\XHeA}(p_1,p_2,k_1,k_2,k) = 
\Bigl( \CxiETA - \XIeta \Bigr)
\frac{\langle p_1(P_2-K) \KH p_1\rangle^*}{\Kpo\Kpt},}\quad
\nl[.5em]
\lefteqn{A_{+,\tau_1,\tau_2,+}^{\XHtA}(p_1,p_2,k_1,k_2,k) = 
\langle p_2\KH p_1\rangle}\quad\nl*
&&{}\times\biggl\{\frac{\KKtKoK}{4\MKoK\MKtK}
\Bigl( \CxiETA-\XIeta \Bigr) - \CKxi\CKETA
\biggl[ \frac{1}{2\MKoK}+\frac{1}{2\MKtK} \biggr]\biggr\},
\qquad
\nl[.5em]
\lefteqn{A_{+,\tau_1,\tau_2,+}^{\VtHeA}(p_1,p_2,k_1,k_2,k) =
\frac{1}{\Kpo\Kpt}
\biggl[
\gVtm\langle \eta'(P_2-K)p_1\rangle\Bigl(2\Mt\poeta+\langle \xi\KH p_1\rangle\Bigr)
}\quad
\nl*&&{}
+\gVtp\poXI\Bigl(2\Mt\langle \xi(P_2-K)p_1\rangle+\langle p_1(P_2-K)\KH\eta\rangle^*\Bigr)
\biggr]
,\nl[0.5em]
\lefteqn{A_{+,\tau_1,\tau_2,+}^{\VtHtA}(p_1,p_2,k_1,k_2,k) =
}\quad\nl*
&=&\frac{1}{2\MKoK}
P_\Pt(k_1+\kH) \,
\left[
\langle k \KH K_1k\rangle P_\Pt(k_1+\kH+k) \,
-\frac{\KKtKoK}{2\MKtK}\right]
\nl*&&{}\times
\left[\gVtp\poXI\left(2\Mt\Cptxi-\langle p_2\KH\eta\rangle\right)
+\gVtm\CptETA\left(2\Mt\poeta+\langle\xi\KH p_1\rangle\right)\right]
\nl&&{}
-\frac{\CKxi}{2\MKoK} P_\Pt(k_1+\kH+k) \,
\left[2\gVtp\Mt\poXI\CKpt-\gVtm\CptETA\langle k \KH p_1\rangle\right]
\nl&&{}
+\frac{\CKETA\CKpt}{2\MKtK} P_\Pt(k_1+\kH) \,
\gVtm\left[2\Mt\poeta+\langle\xi \KH p_1\rangle\right]
\nl*&&{}
-\CKpt\poXI P_\Pt(k_1+\kH) \, P_\Pt(k_1+\kH+k) \,
\gVtp\left[2\Mt\CKxi-\langle k \KH\eta\rangle\right],
\nl[.5em]
\lefteqn{A_{+,\tau_1,\tau_2,+}^{\VtbarHeA}(p_1,p_2,k_1,k_2,k) =
\frac{1}{\Kpo\Kpt} 
\biggl[
\gVtp\langle \xi(P_2-K)p_1\rangle\Bigl(2\Mt\poXI+\langle \eta'\KH p_1\rangle\Bigr)
}\quad
\nl*&&{}
+\gVtm\poeta\Bigl(2\Mt\langle \eta'(P_2-K)p_1\rangle+\langle p_1(P_2-K)\KH\xi'\rangle^*\Bigr)
\biggr]
,\nl[.5em]
\lefteqn{A_{+,\tau_1,\tau_2,+}^{\VtbarHtA}(p_1,p_2,k_1,k_2,k)
  =}\quad\nl*
&=&{}
-\frac{1}{2\MKtK} P_\Pt(k_2+\kH) \,
\left[
\langle k \KH K_2k\rangle P_\Pt(k_2+\kH+k) \,
+\frac{\KKtKoK}{2\MKoK}\right]
\nl*&&{}\times
\left[\gVtp\Cptxi\left(2\Mt\poXI+\langle \eta'\KH p_1\rangle\right)
+\gVtm\poeta\left(2\Mt\CptETA-\langle p_2\KH\xi'\rangle\right)\right]
\nl&&{}
+\frac{\CKETA}{2\MKtK} P_\Pt(k_2+\kH+k) \,
\left[2\gVtm\Mt\poeta\CKpt-\gVtp\Cptxi\langle k \KH
  p_1 \rangle\right]
\nl&&{}
-\frac{\CKxi\CKpt}{2\MKoK} P_\Pt(k_2+\kH) \,
\gVtp\left[2\Mt\poXI+\langle\eta'\KH p_1 \rangle\right]
\nl*&&{}
+\CKpt\poeta P_\Pt(k_2+\kH) \, P_\Pt(k_2+\kH+k) \,
\gVtm\left[2\Mt\CKETA-\langle k \KH\xi'\rangle\right]
\eeqar
and the symmetry relations
\beqar\label{eq:symend}
A_{-\si,\tau_1,\tau_2,\la}^{\ldots\Pt}(p_1,p_2,k_1,k_2,k) &=&
A_{\si\tau_1\tau_2\la}^{\ldots\Pt}(p_2,p_1,k_1,k_2,k),\nl
A_{-\si,\tau_1,\tau_2,\la}^{\ldots\Pe}(p_1,p_2,k_1,k_2,k) &=&
-A_{\si\tau_1\tau_2\la}^{\ldots\Pe}(p_2,p_1,k_1,k_2,k),\nl
A_{\si,-\tau_2,-\tau_1,-\la}^{Bf}(p_1,p_2,k_1,k_2,k) &=&
\sgn(\tau_1
\tau_2)A_{\si\tau_1\tau_2\la}^{Bf}(p_2,p_1,k_2,k_1,k)^*,\quad
B=\PZ,\chi,\quad f=\Pe,\Pt,
\nl
A_{\si,-\tau_2,-\tau_1,-\la}^{V\Pt f}(p_1,p_2,k_1,k_2,k) &=&
\sgn(\tau_1
\tau_2)A_{\si\tau_1\tau_2\la}^{V\Ptbar f}(p_2,p_1,k_2,k_1,k)^*,\quad
 f=\Pe,\Pt,
\nl
A_{\si,-\tau_2,-\tau_1,-\la}^{V\Ptbar f}(p_1,p_2,k_1,k_2,k) &=&
\sgn(\tau_1
\tau_2)A_{\si\tau_1\tau_2\la}^{V\Pt f}(p_2,p_1,k_2,k_1,k)^*,\quad
 f=\Pe,\Pt.
\eeqar
Note that interchanging $k_1\leftrightarrow k_2$ means also
interchanging $\eta\leftrightarrow \xi',\xi\leftrightarrow \eta'$.
Besides \refeq{eq:auxkh} we used the following auxiliary quantities 
in the previous definitions:
\beqar
\langle k K_2 K_1 k \rangle &=&
k_{\dot A} K_2^{\dot AB} K_{1,B\dot C} k^{\dot C} = \sum_{i,j=1,2}
\langle k \rho'_i \rangle^* \langle \rho_j \rho'_i \rangle
\langle \rho_j k \rangle^*,\nl
\langle k \KH K_1 k \rangle &=&
k_{\dot A} \KH^{\dot AB} K_{1,B\dot C} k^{\dot C} = \sum_{i,j=1,2}
\langle k \phi_i \rangle^* \langle \rho_j \phi_i \rangle
\langle \rho_j k \rangle^*,\nl
\langle k\KH K_2 k \rangle &=&
k_{\dot A} \KH^{\dot AB} K_{2,B\dot C} k^{\dot C} = \sum_{i,j=1,2}
\langle k \phi_i \rangle^* \langle \rho'_j \phi_i \rangle
\langle \rho'_j k \rangle^*,\nl
\langle \psi' (P_2-K)\KH \psi \rangle &=&
\psi'_{\dot A} P_2^{\dot AB} K_{3,B\dot C} \psi^{\dot C} 
-\psi'_{\dot A} K^{\dot AB} K_{3,B\dot C} \psi^{\dot C} 
\nl*&=& \sum_{j=1,2}\left(
\langle \psi' p_{2} \rangle^* \langle \phi_j  p_{2}\rangle
\langle \phi_j \psi \rangle^*
-\langle \psi' k \rangle^* \langle \phi_j  k\rangle
\langle \phi_j \psi \rangle^*\right).
\label{eq:kKtKok}
\eeqar
The Weyl spinor $k_A$ and the momentum matrix $K_{\dot AB}$ of the
photon are defined from the photon momentum $k$ analogously to those
of the massless electron in \refeqs{eq:electronspinor} and
\refeqf{eq:electronmomentummatrix}.

The contribution $\sigma_\gamma$ of the radiative process to the cross
section is given by
\beqar\label{eq:hbcs}
\sigma_\gamma &=& \frac{1}{2s} \int \rd\Phi_\gamma \,
 \sum_{\si=\pm\frac12} \frac{1}{4}
(1+2P_-\si)(1-2P_+\si) \, |\M_\ga^{\si}|^2
\eeqar
with
\beqar
|\M_\ga^{\si}|^2&=& \NCt\sum_{\lambda=\pm1} \,
\sum_{\tau_1=\pm\frac12} \,\sum_{\tau_2=\pm\frac12} \,
|\M^{\si\tau_1\tau_2\la}_\gamma|^2,
\eeqar
and the phase-space integral 
\beq
\int \rd\Phi_\gamma =
\int\frac{\rd^3 {\bf k}}{(2\pi)^3 2k^0} \,
\left( \prod_{i=1}^3 \int\frac{\rd^3 {\bf k}_i}{(2\pi)^3 2k_i^0} \right)\,
(2\pi)^4 \delta\Biggl(p_1+p_2-k-\sum_{j=1}^3 k_j\Biggr).
\label{eq:dPSg}
\eeq

The spin-averaged squared
matrix element obtained from Eqs.~\refeq{eq:MEttHg}--\refeq{eq:symend}
has been successfully checked against the result obtained with the 
package {\sl Madgraph} \cite{Stelzer:1994ta} numerically.

\subsubsection{Treatment of soft and collinear singularities}

Without soft and collinear regulators the phase-space integral
\refeq{eq:hbcs} diverges in the soft ($k_0\to 0$) and collinear ($p_a
k\to 0$) phase-space regions.  In the following we describe two
procedures of treating soft and collinear photon emission: one is
based on a subtraction method, the other on phase-space slicing. In
both cases soft and collinear singularities are regularized by an
infinitesimal photon mass and a small electron mass, respectively.

Comparing both methods we find numerical agreement for total cross
sections and distributions within integration errors. The results
presented in this paper were obtained with the subtraction method, and
the statistical error is typically $0.1\%$ or below for the total
cross section.

\paragraph{The dipole subtraction approach}

The idea of so-called subtraction methods is to subtract a simple
auxiliary function from the singular integrand of the bremsstrahlung
integral and to add this contribution back again after partial
analytic integration.  This auxiliary function, denoted $|\M_\sub|^2$
in the following, has to be chosen in such a way that it cancels all
singularities of the original integrand, which is $|\M_\ga|^2$ in our
case, so that the phase-space integration of the difference can be
performed numerically, even over the singular regions of the original
integrand.  In this difference, $\M_\ga$ can be evaluated without
regulators for soft or collinear singularities, i.e.\ we can make use
of the results of the previous section.  The auxiliary function has to
be simple enough so that it can be integrated over the singular
regions analytically, when the subtracted contribution is added again.
This part contains the singular contributions and requires regulators,
i.e.\ photon and electron masses have to be reintroduced there.
Specifically, we have applied the {\it dipole subtraction formalism},
which is a process-independent approach that was first proposed
\cite{Catani:1996jh} within QCD for massless unpolarized partons and
subsequently generalized to photon radiation of massive polarized
fermions in \citere{Dittmaier:2000mb}.  In order to keep the
description of the method transparent, we describe only the basic
structure of the individual terms explicitly and refer to
\citere{Dittmaier:2000mb} for details.

In the dipole subtraction formalism the subtraction functions are
constructed from contributions related to ordered pairs of charged
fermions.  These two fermions are called {\it emitter} and {\it
  spectator}, respectively, since by construction only the kinematics
of the emitter leads to collinear singularities. In the following, we
consider only the case for unpolarized final-state fermions.  The
subtraction function receives four kinds of contributions depending
whether the emitter (first index) and spectator (second index) are
chosen from initial (labelled by $a,b$) or final state (denoted by
$i,j$):
\beqar
|\M^\si_\sub|^2 &=& \sum_{a,b=1,2 \atop a\ne b}|\M^\si_{\sub,ab}|^2 
+\sum_{a=1,2 \atop i=1,2}\left(|\M^\si_{\sub,ai}|^2 
+|\M^\si_{\sub,ia}|^2 \right)
+\sum_{i,j=1,2 \atop i\ne j} |\M^\si_{\sub,ij}|^2  \qquad
\eeqar
with the contributions
\beqar
|\M^\si_{\sub,ab}|^2 &=&
e^2 \gsub_{ab,+}(p_a,p_b,k)
\left|\M^\si_0(\tilde p_1,\tilde p_2,\tilde k_1,\tilde k_2,\tilde \kH)\right|^2,
\label{eq:msubab}
\nl
|\M^\si_{\sub,ai}|^2 &=&
(-1)^{a+i+1}\Qt e^2 \gsub_{ai,+}(p_a,k_i,k)
\left|\M^\si_0(\tilde p_1,\tilde p_2,\tilde k_1,\tilde k_2,\tilde \kH)\right|^2,
\label{eq:msubai}
\nl
|\M^\si_{\sub,ia}|^2 &=&
(-1)^{a+i+1}\Qt e^2 \gsub_{ia}(k_i,p_a,k)
\left|\M^\si_0(\tilde p_1,\tilde p_2,\tilde k_1,\tilde k_2,\tilde \kH)\right|^2,
\label{eq:msubia}
\nl
|\M^\si_{\sub,ij}|^2 &=&
\Qt^2 e^2 \gsub_{ij}(k_i,k_j,k)
\left|\M^\si_0(\tilde p_1,\tilde p_2,\tilde k_1,\tilde k_2,\tilde \kH)\right|^2.
\label{eq:msubij}
\eeqar
The unpolarized dipole functions for final-state emitters read
\beqar
\gsub_{ia}(k_i,p_a,k)&=&\gsub_{ia,+}(k_i,p_a,k)+\gsub_{ia,-}(k_i,p_a,k),\nl
\gsub_{ij}(k_i,k_j,k)&=&\gsub_{ij,+}(k_i,k_j,k)+\gsub_{ij,-}(k_i,k_j,k).
\eeqar
The dipole functions $\gsub_{\ldots}$ are defined in Eqs.\ (3.22),
(A.1), and (4.4) of \citere{Dittmaier:2000mb}.  Note that the
spin-flip dipole functions $\gsub_{ab,-}$ and $\gsub_{ai,-}$ vanish in
the case of small fermion masses, i.e.\ $\Me \to 0$.  The arguments
$\tilde p_a,\tilde k_i$ of the tree-level matrix elements $\M^{\si}_0$
in subtraction functions depend on the momenta $p_a,k_i,k$.  The
mapping from $p_a,k_i,k$ into $\tilde p_a,\tilde k_i$ is defined by
Eqs.\ (3.25) and (3.26) of \citere{Dittmaier:2000mb} for
$\M^\si_{\sub,ab}$, by Eq.\ (3.12) for $\M^\si_{\sub,ai}$ and
$\M^\si_{\sub,ia}$, and by Eq.\ (4.5) for $\M^\si_{\sub,ij}$.

The subtracted contribution can be integrated over the (singular)
photonic degrees of freedom up to a remaining convolution over $x$
($=x_{ab},x_{ia}$).  In this integration the regulators $m_\gamma$ and
$\Me$ must be retained, and the soft and collinear singularities
appear as logarithms in these mass regulators.  This non-trivial step
has, however, to be done only once and for all, and the needed results
can be found in \citere{Dittmaier:2000mb}.  The integrated subtraction
function reads
\beqar
\lefteqn{\sigma_{\sub}(p_1,p_2,P_-,P_+)=}\quad
\nn \\ &=&
\frac{\alpha}{2\pi} \int_0^1\rd x\, \Biggl\{
\sum_{\tau=\pm} \, \sum_{a=1,2} \,
\int\rd\sigma_0^{(a,\tau)}(x)\,
\Biggl[\sum_{b=1,2 \atop b\ne a}\cGsub_{ab,\tau}(s,x)
+\sum_{i=1,2}(-1)^{a+i+1}\Qt\cGsub_{ai,\tau}(t_{ai},x)
\Biggr]_+
\nn\\ && {}
\qquad +\sum_{a=1,2 \atop i=1,2}
\int\rd\sigma_0^{(a,+)}(x) \,(-1)^{a+i+1}\Qt
\left[\cGsub_{ia}(t_{ai},x)\right]_+
\Biggr\}
\nn\\ && {}
+ \frac{\alpha}{2\pi} \Biggl\{
\sum_{\tau=\pm} \, \sum_{a=1,2} \,
\int\rd\sigma_0^{(a,\tau)}(1)\,
\Bigg[\sum_{b=1,2 \atop b\ne a}\Gsub_{ab,\tau}(s)
+\sum_{i=1,2}(-1)^{a+i+1}\Qt\Gsub_{ai,\tau}(t_{ai}) \Bigg]
\nn\\ && {}
\qquad + \int\rd\sigma_0 \,
\Bigg[ \sum_{a=1,2 \atop i=1,2}(-1)^{a+i+1}\Qt\Gsub_{ia}(t_{ai})
+\sum_{i,j=1,2 \atop i\ne j}\Qt^2\Gsub_{ij}(s_{ij}) \Bigg]
\Biggr\}
\eeqar
with the modified lowest-order cross sections
\beq\label{eq:boostedcs}
\rd\sigma_0^{(1,\tau)}(x) = \rd\sigma_0(xp_1,p_2,\tau P_-,P_+), \qquad
\rd\sigma_0^{(2,\tau)}(x) = \rd\sigma_0(p_1,xp_2,P_-,\tau P_+)
\eeq
and the unpolarized distributions $\cGsub$ and end-point contributions
$\Gsub$, 
\beqar
\cGsub_{ia}(t_{ai},x)&=&\cGsub_{ia,+}(t_{ai},x)+\cGsub_{ia,-}(t_{ai},x), \nl
\Gsub_{ia}(t_{ai})&=&\Gsub_{ia,+}(t_{ai})+\Gsub_{ia,-}(t_{ai}),\nl
\Gsub_{ij}(s_{ij})&=&\Gsub_{ij,+}(s_{ij})+\Gsub_{ij,-}(s_{ij}).
\eeqar
In our case, the relevant definitions for $\cGsub_{\ldots}$ and
$\Gsub_{\ldots}$ are given in Eqs.\ (3.32), (3.33), (A.2), (A.4), and
(4.10) of \citere{Dittmaier:2000mb}.  The $[\dots]_+$ prescription is
defined as usual,
\beq
\int_0^1\rd x\, \Bigl[f(x)\Bigr]_+ g(x) \equiv
\int_0^1\rd x\, f(x) \left[g(x)-g(1)\right].
\eeq
In summary, the phase-space integral \refeq{eq:hbcs} in the dipole
subtraction approach reads
\beq
\sigma_\gamma = \frac{1}{2s} \int \rd\Phi_\gamma \,
\sum_{\si=\pm\frac{1}{2}} \frac{1}{4} 
(1+2P_-\si)(1-2P_+\si) \, 
\left[|\M^{\si}_\gamma|^2
-|\M^\si_\sub|^2 \right] \;+\; \sigma_{\sub}.
\eeq

\paragraph{The phase-space-slicing approach}

The idea of the phase-space-slicing method is to divide the
bremsstrahlung phase space into singular and non-singular regions,
then to evaluate the singular regions analytically and to perform an
explicit cancellation of the arising soft and collinear singularities
against their counterparts in the virtual corrections. The finite
remainder can be evaluated by Monte Carlo techniques.  For the actual
implementation of this well-known procedure we closely follow the
approach of \citere{bo93}.  We divide the four-particle phase space
into soft and collinear regions by introducing the cut-off parameters
$\De E$ and $\De\theta$, respectively, and decompose the real
corrections as
\begin{equation}
\rd\sigma_{\ga} = \rd\sigma_{\soft}+ \rd\sigma_{\coll}+\rd\sigma_{\ga,\finite}.
\end{equation}
Here $\rd\sigma_{\soft}$ describes the contribution of the soft
photons, \ie of photons with energies $k_0 < \Delta E$ in the CM
frame, and $\rd\sigma_{\coll}$ describes real photon radiation outside
the soft-photon region ($k_0>\Delta E$) but collinear to the $\Pe^\pm$
beams.  The collinear region consists of the two disjoint parts
$0<\theta_{\ga}<\De\theta$ and $\pi-\De\theta<\theta_{\ga}<\pi$, where
$\theta_{\ga}$ is the polar angle of the emitted photon in the CM
frame. The remaining part, which is free of singularities, is denoted
by $\rd\sigma_{\ga,\finite}$.

In the soft and collinear regions, the squared matrix element
$|\M^\si_{\ga}|^2$ factorizes into the leading-order squared matrix
element $|\M^\si_0|^2$ and a soft or collinear factor.  Also the
four-particle phase space factorizes into a three-particle phase space
and a soft or collinear part, so that the integration over the
singular part of the photon phase space can be performed analytically.

In the soft-photon region, we apply the soft-photon approximation to
$|\M^\si_{\ga}|^2$, \ie the photon four-momentum $k$ is omitted
everywhere but in the IR-singular propagators. In this region
$\rd\sigma_{\ga}$ can be written as \cite{Denner:1993kt,Yennie:ad}
\beq
\rd\sigma_{\soft} = - \rd\sigma_0
\frac{\alpha}{4\pi^2} 
\int_{k_0< \Delta E \atop k_0^2 = |{\bf k}|^2 +m_\ga^2}
\frac{\rd^3 {\bf k}}{k_0}
\left( \frac{p_1^{\mu}}{p_1 k} - \frac{p_2^{\mu}}{p_2 k}
  + \Qt \frac{k_1^\mu}{k_1 k} - \Qt \frac{k_2^\mu}{k_2 k}
\right)^2.
\eeq
This can be decomposed into an ISR part, a final-state-radiation (FSR)
part (proportional to $Q_\Pt^2$) and the ISR--FSR-interference part
(proportional to $Q_\Pt$),
\beq
\rd\sigma_{\soft}= \dsigma_{\soft,\mathrm{ISR}} + \dsigma_{\soft,\mathrm{FSR}}
   + \dsigma_{\soft,\mathrm{int}}.
\eeq

The explicit expressions for the soft-photon integrals can be found in
\citeres{Denner:1993kt,'tHooft:1979xw}.  For our purpose it is
sufficient to keep the electron mass only as regulator for the
collinear singularities. In this limit we obtain for the ISR part
\beq
\rd\sigma_{\soft,\mathrm{ISR}} = - \rd\sigma_0 \, \frac{\alpha}{\pi}
\left\{ 2 \ln\left(\frac{2\Delta E}{m_\ga}\right)
\, \left[1-\ln\left(\frac{s}{\Me^2}\right)\right]
-\ln\left(\frac{s}{\Me^2}\right)
+\frac{1}{2}\ln^2\left(\frac{s}{\Me^2}\right)+
\frac{\pi^2}{3}
\right\}.
\label{eq:si_soft}
\eeq
The FSR part reads
\beq
\dsigma_{\soft,\mathrm{FSR}} = - \rd\sigma_0 \frac{\alpha}{4\pi^2} \Qt^2
\left[ I_{\Pt\Pt} + I_{\Ptbar\Ptbar} - 2 I_{\Pt\Ptbar} \right]
\eeq
with the massive integrals $I_{ii}$ and $I_{ij}$ from
\citeres{Denner:1993kt,'tHooft:1979xw}.  Finally, the interference
contribution is given by
\beq
\dsigma_{\soft,\mathrm{int}} = - \rd\sigma_0 \frac{\alpha}{2\pi^2} \Qt
\left[ I_{\Pem\Pt} - I_{\Pem\Ptbar} - I_{\Pep\Pt} + I_{\Pep\Ptbar} \right]
\eeq
in terms of the soft integral with a light particle $a$ and a massive
particle $j$,
\beqar
I_{aj} &=& 2\pi \left[ 
  \Ln{\frac{2\Delta E}{m_\gamma}} \Ln{\frac{(2p_ap_j)^2}{m_a^2\,m_j^2}}
  - \frac{1}{4} \Ln[2]{\frac{m_j^2}{(p_{j}^{0} + |{\bp}_j|)^2}} 
  - \frac{1}{4} \Ln[2]{\frac{m_a^2}{(2p_{a}^{0})^2}} 
  - \frac{\pi^2}{6} \right. \nl
&& \qquad \left.
  - \Dilog{1 - \frac{p_{a}^{0} \, m_j^2}{(p_{j}^{0} + |\bp_j|) \: p_ap_j} }
  - \Dilog{1 - \frac{p_{a}^{0}(p_{j}^{0} + |\bp_j|) }{ p_ap_j} }
\right].
\eeqar

In the collinear region, we consider an incoming $\Pe^\mp$ with
momentum $p_a$ being split into a collinear photon and an $\Pe^\mp$
with the resulting momentum $xp_a$ after photon radiation.  The cross
section for hard photon radiation ($k_0>\Delta E$) in this region
reads
\beqar
\sigma_{\coll}(p_1,p_2,P_-,P_+) &=&
\frac{\alpha}{2\pi} \int_0^{1-2\De E/\sqrt{s}}\rd x\,
\sum_{\tau=\pm} \, \cGcoll_{\tau}(s,x)\Biggl[ 
\int\rd\sigma_0^{(1,\tau)}(x)+\int\rd\sigma_0^{(2,\tau)}(x) \Biggr],
\nn\\
\label{eq:si_collin}
\eeqar
where \refeq{eq:boostedcs} is used, and
\beqar
\cGcoll_+(s,x) &=&
\frac{1+x^2}{1-x}
\left[\ln\biggl(\frac{s\De\theta^2}{4\Me^2}\biggr)-1\right], \qquad
\cGcoll_-(s,x) = 1-x.
\eeqar

Finally, the finite cross section $\rd\sigma_{\ga,\finite}$ is defined
by imposing cuts on the bremsstrahlung phase space \refeq{eq:dPSg},
\ie a photon-energy cut, $k_0>\Delta E$, and a cut on the angles
between the photon and the beams,
$\De\theta<\theta_{\ga}<\pi-\De\theta$.

For the numerical evaluation the values $\Delta E/\sqrt{s} = 10^{-6}$
and $\Delta\theta = 10^{-3}$ have been chosen.  For $\sqrt{s}=500\GeV$
and $\MH=140\GeV$ as well as for $\sqrt{s}=1\TeV$ and $\MH=115\GeV$,
we have checked that the cross section changes only within integration
errors when varying $\Delta E$ within $10^{-5} \GeV < \Delta E <
10^{-2} \GeV$ for fixed $\Delta\theta = 10^{-3}$ and when varying
$\Delta\theta$ within $10^{-5} < \Delta\theta < 10^{-2}$ for fixed
$\Delta E = 10^{-3} \GeV$.

\subsection{QCD corrections}
\label{se:QCD}

The ${\cal O}(\alphas)$ QCD corrections can be obtained from the
photonic FSR part of the $\Oa$ correction, \ie the part proportional
to $Q_\Pt^2$, by the replacement
\beq
Q_\Pt^2\alpha \to C_{\mathrm{F}}\alphas(\mu^2)=\frac{4}{3}\alphas(\mu^2),
\eeq
where $Q_\Pt=2/3$ is the relative charge of the top quark.  The real
part of the FSR corrections results from diagrams where a photon is
emitted from top quarks. The corresponding matrix elements
$\M_\gamma^{\ldots\Pt}$ are given in Eqs.\ \refeq{eq:ZHtA},
\refeq{eq:XHtA}, \refeq{eq:VtHtA}, and \refeq{eq:VtbarHtA}.  Following
\citere{Dittmaier:1998dz}, the QCD renormalization scale $\mu$ is set
to the CM energy $\sqrt{s}$, and the running of the strong coupling is
evaluated at the two-loop level ($\overline{\mathrm{MS}}$ scheme) with
five active flavours, normalized by $\alphas(\MZ^2)$ as given in
\refeq{eq:SMpar}.  For $\sqrt{s}=500\GeV$, $800\GeV$, and $1000\GeV$
the resulting values for the strong coupling are given by
$\alphas(\MZ^2)= 0.09349$, $0.08857$, and $0.08642$, respectively.

\subsection{Initial-state radiation beyond ${\cal O}(\alpha)$}
\label{se:ISR}

The emission of photons collinear to the incoming electrons or
positrons leads to corrections that are enhanced by large logarithms
of the form $\ln(\Me^2/s)$.  In order to achieve an accuracy at the
few $0.1\%$ level, the corresponding higher-order contributions, i.e.\ 
contributions beyond $\Oa$, must be taken into account. These are
included in our calculation through a convolution of the leading-order
cross section with leading-logarithmic structure functions precisely
in the same way as described in \citere{Denner:2003yg}.

\subsection{Monte Carlo integration}

The phase-space integration is performed with Monte Carlo techniques
in both computer codes. The first code employs a multi-channel Monte
Carlo generator similar to the one implemented in {\sl RacoonWW}
\cite{Roth:1999kk,Denner:1999gp} and {\sl Lusifer}
\cite{Dittmaier:2002ap}, the second one uses the adaptive
multi-dimensional integration program {\sl VEGAS}
\cite{Lepage:1977sw}.

\section{Numerical results}
\label{se:numres}

\subsection{Input parameters}

For the numerical evaluation we use the following set of SM parameters
\cite{Hagiwara:pw}:
\beq
\begin{array}[b]{lcllcllcl}
\GF & = & 1.16639 \times 10^{-5} \GeV^{-2}, \ \ &
\alpha(0) &=& 1/137.03599976, &
\alphas(\MZ) &=& 0.1172, \\
\MW & = & 80.423\GeV, &
\MZ & = & 91.1876\GeV, \\
\Me & = & 0.510998902\MeV, &
m_\mu &=& 105.658357\MeV,\ \  &
m_\tau &=& 1.77699\GeV, \\
\Mu & = & 66\MeV, &
\Mc & = & 1.2\GeV, &
\Mt & = & 174.3\;\GeV, \\
\Md & = & 66\MeV, &
\Ms & = & 150\MeV, &
\Mb & = & 4.3\GeV,
\end{array}
\label{eq:SMpar}
\eeq
which coincides with the one used in
\citeres{Denner:2003ri,Denner:2003yg}.  Since we employ the on-shell
renormalization scheme, the weak mixing angle is fixed by
\refeq{eq:defcw}. Because quark mixing is irrelevant for the
considered process we use a unit quark-mixing matrix.

We evaluate amplitudes in the so-called $\GF$-scheme, i.e.\ we derive
the electromagnetic coupling $\alpha=e^2/(4\pi)$ from the Fermi
constant $\GF$ according to
\beq\label{eq:defGF}
\alpha_{\GF} = \frac{\sqrt{2}\GF\MW^2\sw^2}{\pi}.
\eeq
This procedure, in particular, absorbs all sizable mass effects of
light fermions other than electrons in the coupling $\alpha_{\GF}$,
and the results are practically independent of the masses of the light
quarks. The masses of the light quarks are adjusted to reproduce the
hadronic contribution to the photonic vacuum polarization of
\citere{Jegerlehner:2001ca}.  In the relative radiative corrections,
we use $\alpha(0)$ as coupling parameter, which is the correct
effective coupling for real photon emission. We do not calculate the
W-boson mass from $\GF$ but use its experimental value as input.

\subsection{Results on total cross sections}

Numerical results for total cross sections including individual
contributions to the radiative corrections have already been presented
in \citere{Denner:2003ri}. The contributions discussed there were the
\textit{photonic} corrections originating from virtual photon exchange
and real photon radiation, the beyond $\Oa$ or \textit{higher-order
  ISR} corrections that are included using the structure-function
approach as described above, and the remaining non-photonic
\textit{weak} corrections. All these have been combined into the
complete \textit{electroweak} corrections.  Furthermore, the
\textit{QCD} corrections have been considered.

\begin{figure}
\centerline{
\includegraphics[width=.5\textwidth,bb=78 415 279 628]{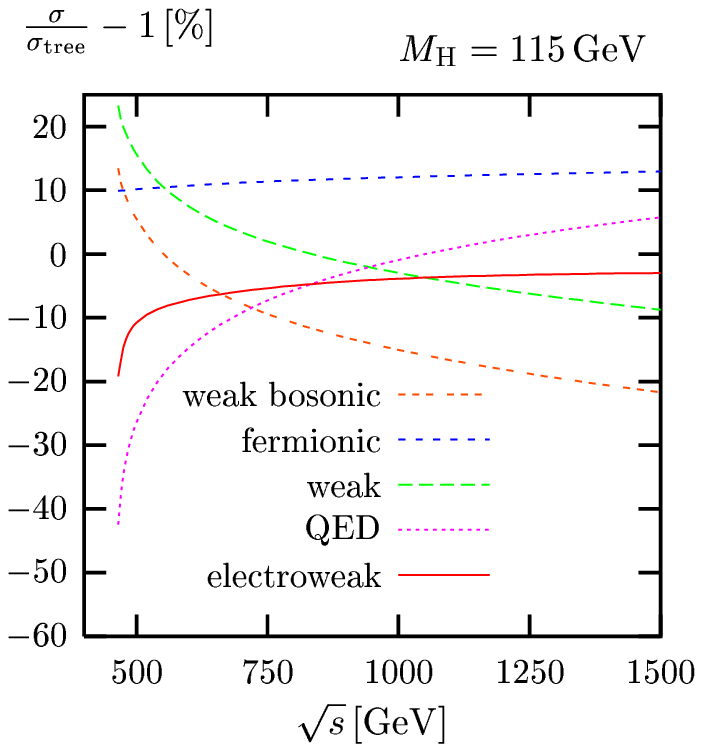}
\includegraphics[width=.5\textwidth,bb=78 415 279 628]{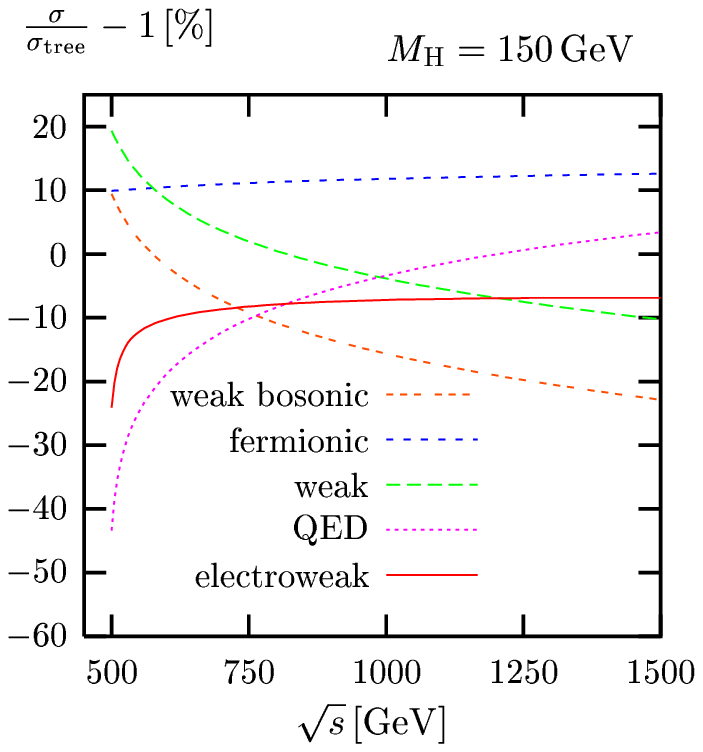}
}
\caption{Relative corrections for $\MH = 115 \GeV$ (l.h.s) and 
$\MH = 150 \GeV$ (r.h.s)}
\label{fi:tot-sqrts}
\end{figure}

Here, we complement this by a splitting of the \textit{weak}
corrections into the purely \textit{fermionic} loops together with the
fermionic part of the counterterms and the remaining \textit{weak
  bosonic} contributions.  Instead of the separate \textit{photonic}
and \textit{higher-order ISR} contributions considered in
\citere{Denner:2003ri}, we only look at their sum, denoted by {\em
  QED}. The corresponding relative corrections are shown as a function
of the CM energy in \reffi{fi:tot-sqrts} for Higgs-boson masses of
$\MH=115\GeV$ and $150\GeV$.

The fermionic corrections are about $+10\%$ and depend only weakly on
the CM energy. While the weak bosonic contribution is also around
$+10\%$ close to threshold, it falls off rapidly with increasing CM
energies, eventually reaching about $-20\%$ at an energy of $1.5\TeV$.
Therefore, for energies above $600\GeV$, the fermionic and weak
bosonic contributions partially cancel when combined to form the
complete weak corrections. The QED corrections are about $-40\%$ at
threshold rising to a few per cent at $1.5\TeV$. In the electroweak
corrections, both QED and weak contributions partially compensate each
other, yielding relative corrections of about $-20\%$ at threshold,
and $-3\%$ and $-7\%$ for $\MH=115\GeV$ and $150\GeV$, respectively,
at $1.5\TeV$.

\subsection{Results on distributions}

\begin{figure}
\centerline{
\includegraphics[width=.5\textwidth,bb=78 415 279 628]{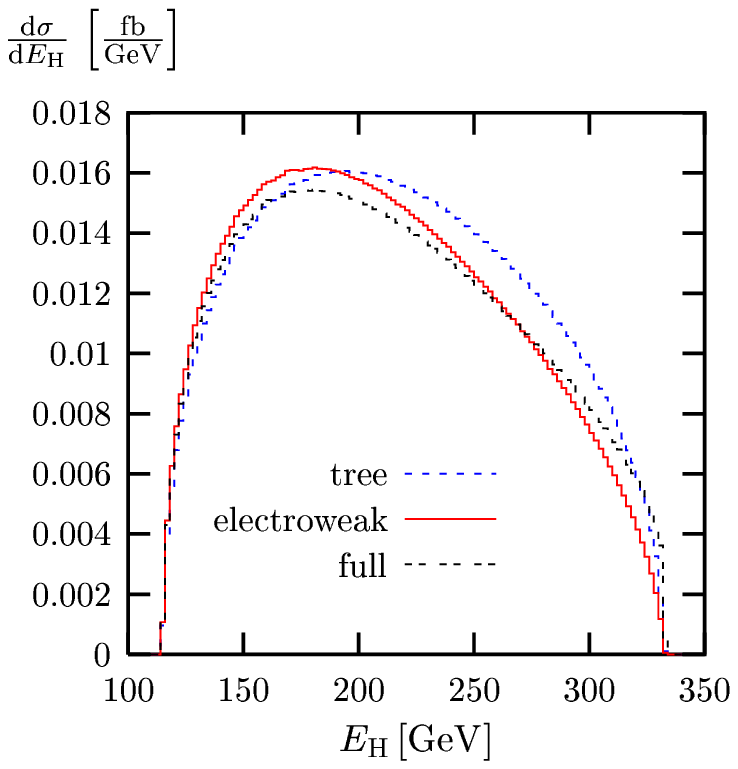}
\includegraphics[width=.5\textwidth,bb=78 415 279 628]{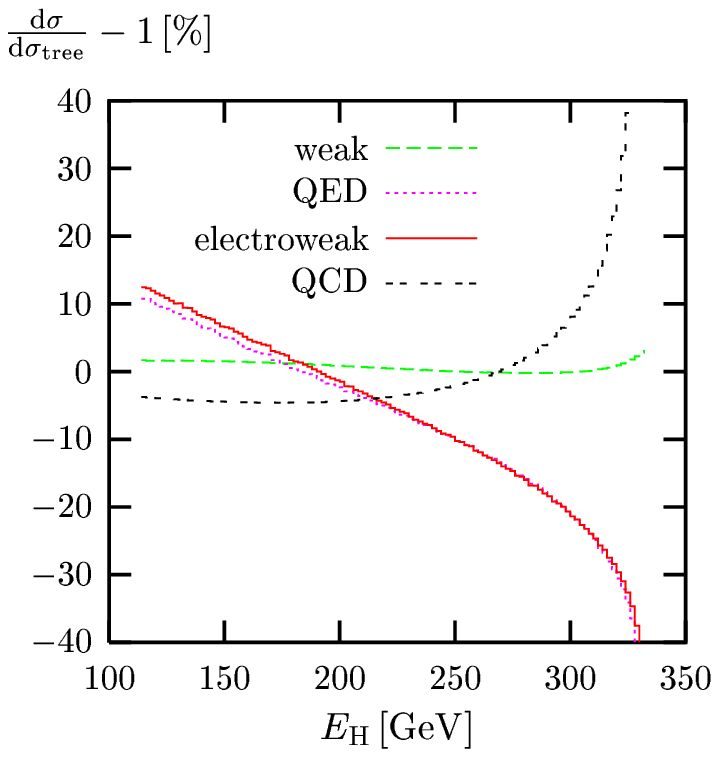}
}
\vspace*{1em}
\centerline{
\includegraphics[width=.5\textwidth,bb=78 415 279 628]{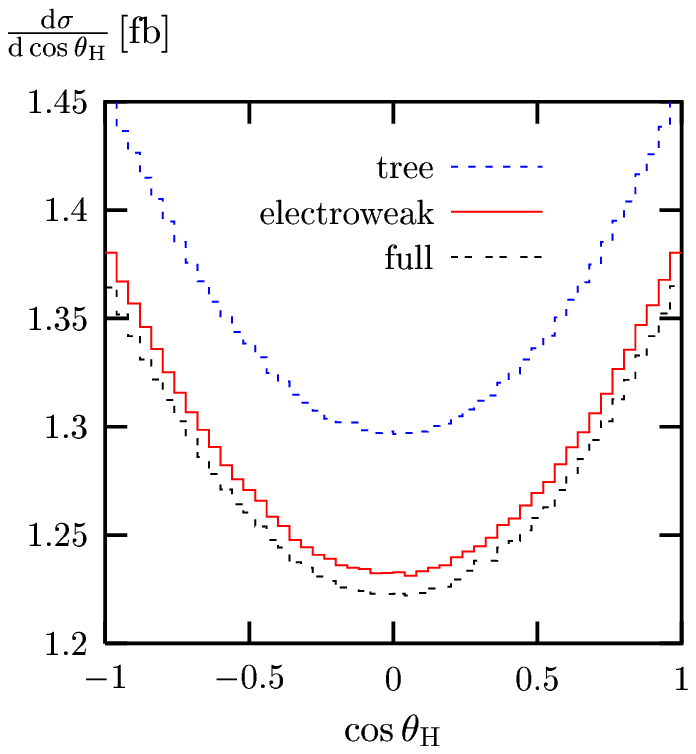}
\includegraphics[width=.5\textwidth,bb=78 415 279 628]{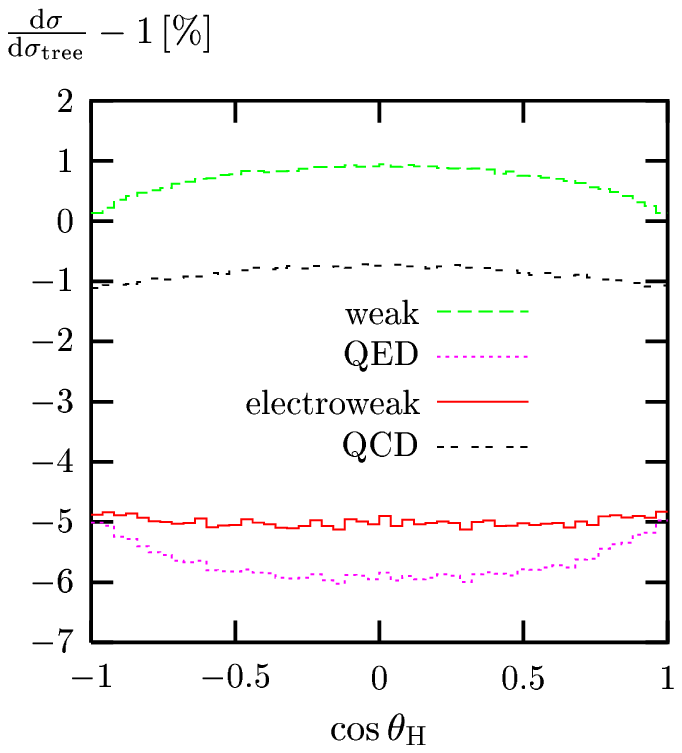}
}
\caption{Distribution in the Higgs-boson energy $E_\PH$ (upper panels)
and the Higgs-boson polar angle $\theta_\PH$ (lower panels) and corresponding
relative corrections for $\sqrt{s}=800\GeV$ and $\MH=115\GeV$}
\label{fi:dist-higgs}
\end{figure}

The distributions in the Higgs-boson energy $E_\PH$ and polar angle
$\theta_\PH$ are shown in \reffi{fi:dist-higgs} for $\MH=115\GeV$ and
$\sqrt{s}=800\GeV$, where the cross section is nearly maximal for
small Higgs-boson masses.
All energies and angles used in the distributions here and below are
defined in the CM frame. The curves denoted with \textit{electroweak}
contain only the electroweak corrections and (in case of the absolute
distributions) the lowest-order contribution, whereas in the curve
labelled \textit{full} the QCD corrections are included as well. The
distribution in the Higgs-boson energy is a broad continuum whose
maximum around $190\GeV$ gets slightly shifted down to smaller
energies by the electroweak corrections but is left unchanged by the
QCD corrections.  At the upper end of the $E_\PH$ spectrum the
$\Pt\bar\Pt$ system becomes non-relativistic, and the QCD correction
is dominated by the Coulomb singularity
\cite{Dittmaier:1998dz,Dittmaier:2000tc,Dittmaier:mg} induced by
soft-gluon exchange in the $\Pt\bar\Pt$ system; the rising QCD
correction with increasing $E_\PH$ reflects the transition to this
singular region.  Although photonic FSR shows the same behaviour, this
effect is totally overruled by ISR in the QED corrections. For
increasing $E_\PH$ these corrections become more and more negative
because of the reduced phase space after ISR, which prohibits a
cancellation of large IR-sensitive virtual corrections by the real
corrections.  For the chosen CM energy and Higgs-boson mass, the
purely weak corrections depend only weakly on the Higgs-boson energy
$E_\PH$ and are only at the per-cent level.  Note, however, that the
smallness of the corrections for the chosen setup is mainly due to the
transition from larger positive to negative corrections with
increasing CM energy, as can be seen in \reffi{fi:tot-sqrts}. Thus,
for lower and higher CM energies sizeable weak corrections to the
$E_\PH$ spectrum arise.
 
The distribution in $\cos\theta_\PH$ reaches maxima in the forward and
backward directions at leading order. The inclusion of the radiative
corrections does not sizeably distort this shape, and essentially only
the normalization is changed. However, viewed separately the weak and
photonic contributions show a small dependence on $\theta_\PH$ which
cancels in their sum.

\begin{figure}
\centerline{
\includegraphics[width=.5\textwidth,bb=78 415 279 628]{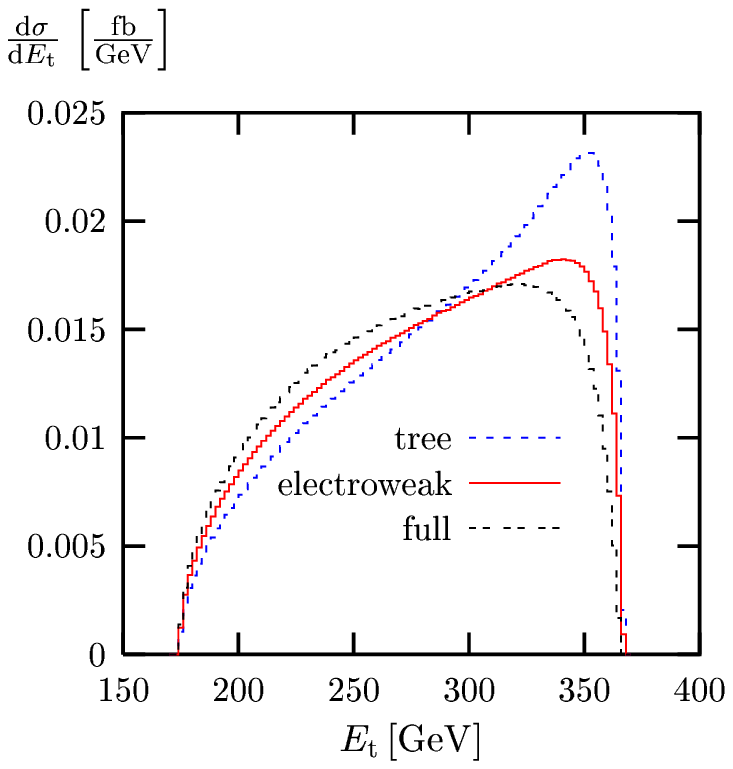}
\includegraphics[width=.5\textwidth,bb=78 415 279 628]{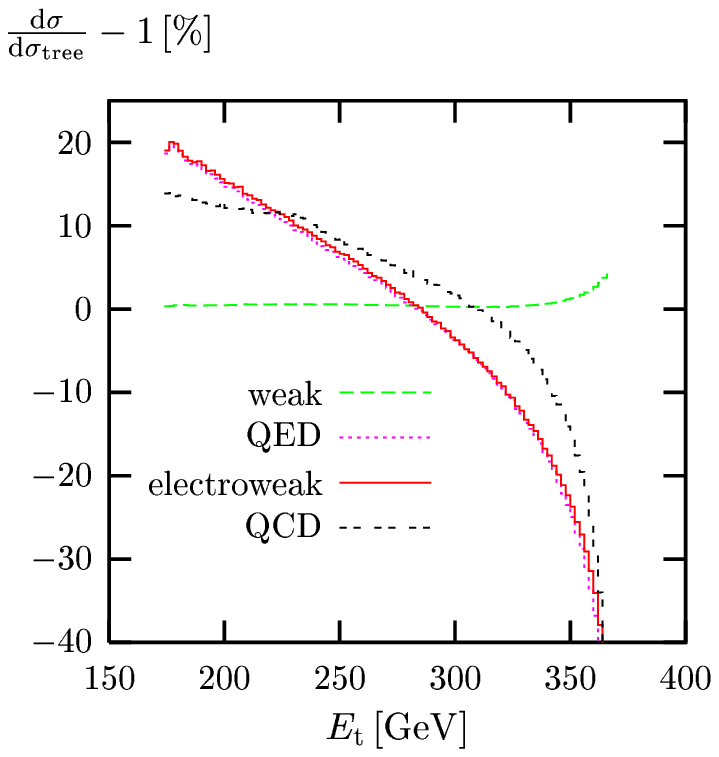}
}
\vspace*{1em}
\centerline{
\includegraphics[width=.5\textwidth,bb=78 415 279 628]{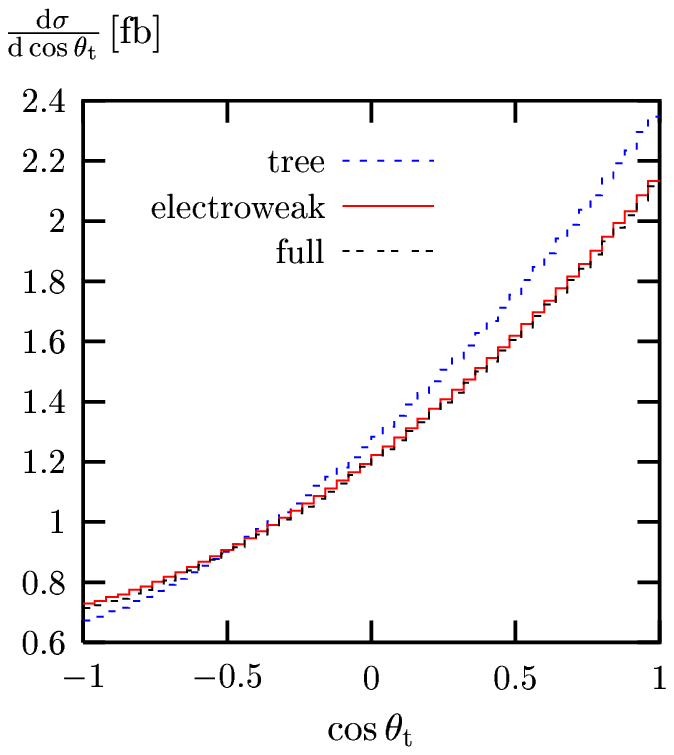}
\includegraphics[width=.5\textwidth,bb=78 415 279 628]{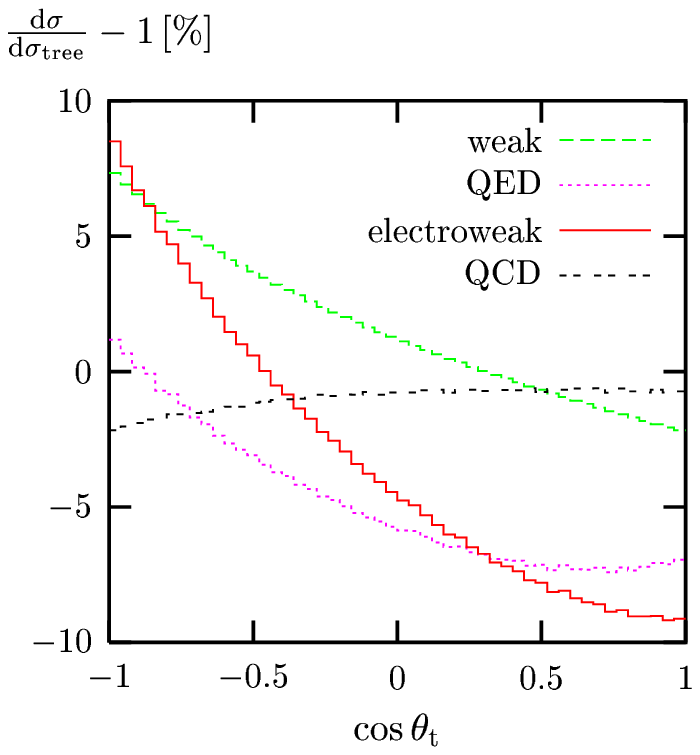}
}
\caption{Distribution in the top-quark energy $E_\Pt$ (upper panels)
and the top-quark polar angle $\theta_\Pt$ (lower panels) and corresponding
relative corrections for $\sqrt{s}=800\GeV$ and $\MH=115\GeV$}
\label{fi:dist-top}
\end{figure}

\reffi{fi:dist-top} shows the distributions in the top-quark energy
$E_\Pt$ and in the cosine of the angle between the top quark and the
incoming electron, $\cos\theta_\Pt$, for $\sqrt{s}=800\GeV$ and
$\MH=115\GeV$.
The $E_\Pt$ distribution has a broad peak at high energies, which is
smeared out by the QED and QCD corrections towards lower energies.
While the weak corrections vary only slowly with $E_\Pt$, QCD and QED
corrections decrease rapidly and reach large negative values for large
$E_\Pt$. This is due to the fact that for large $E_\Pt$, the available
phase space for the emission of a real photon or gluon is restricted
and the large negative virtual corrections can only be partially
compensated.  The distribution in the top-quark production angle is
maximal in the forward ($\Pe^-$) direction. The relative QCD
corrections show only a small angular variation, but the weak and QED
contributions, and hence the electroweak corrections, depend on
$\cos\theta_\Pt$, weakening the maximum somewhat.  Owing to
CP-symmetry the energy distribution of the anti-top quark is identical
to the top-quark distribution. The angular distribution for the
anti-top quark can be obtained from the one of the top quark by the
replacement $\cos\theta_{\Pt} \rightarrow -\cos\theta_{\bar{\Pt}}$.

\begin{figure}
\centerline{
\includegraphics[width=.5\textwidth,bb=78 415 279 628]{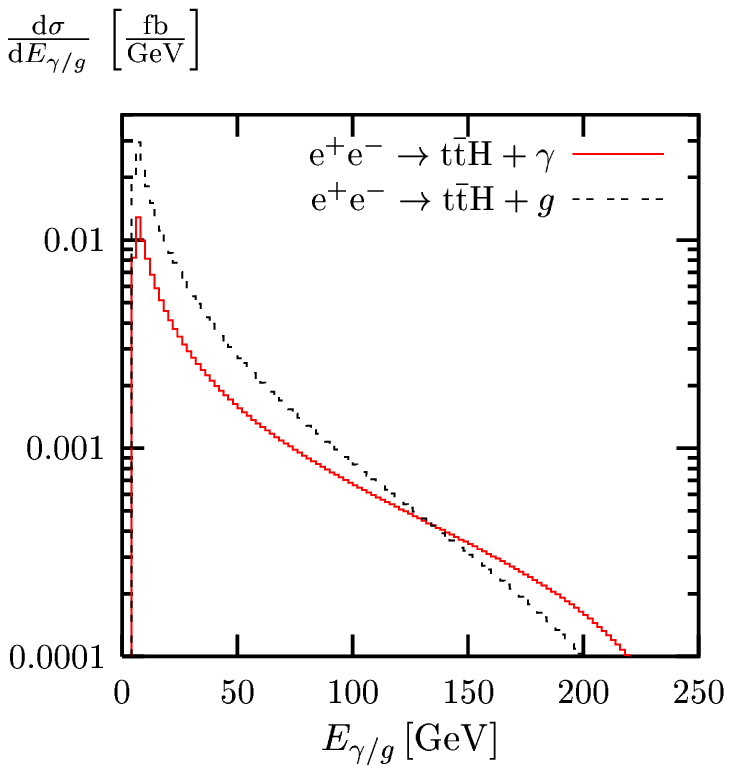}
\includegraphics[width=.5\textwidth,bb=78 415 279 628]{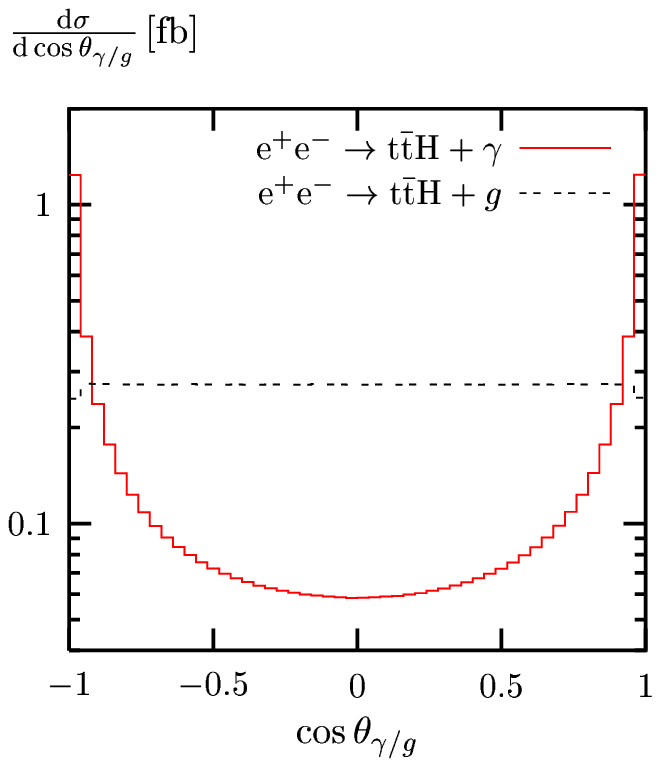}
}
\caption{Distribution in the photon/gluon energy $E_{\gamma/g}$ 
  and in the photon/gluon polar angle $\cos\theta_{\gamma/g}$ in the radiative
  processes $\eetth+\gamma/g$ for $\sqrt{s}=800\GeV$ and $\MH=115\GeV$}
\label{fi:dist-phot}
\end{figure}

Finally, the spectra of the photon energy $E_\gamma$ and of the photon
polar angle $\cos\theta_\gamma$ of the radiative process $\eetth +
\gamma$ are shown in \reffi{fi:dist-phot} together with the
corresponding distributions for the gluon in the process $\eetth + g$,
again for $\sqrt{s}=800\GeV$ and $\MH=115\GeV$.
In order to render the photon/gluon visible, we impose angular and
energy cuts of
\beq 
  \theta(\gamma/g,\mathrm{beam}) > 5^\circ, \qquad E_{\gamma/g}>5\GeV.
\eeq
The photon and gluon energy spectra are dominated by the IR pole in
the region of small energies.  Owing to the collinear singularity of
ISR, the photon angular spectrum shows a strong peaking behaviour in
the forward and backward directions.  In contrast, for the chosen
set-up the gluon radiation does not show a dependence on the gluon
production angle. The reduction in the first and last bins is due to
the angular cut.

\section{Summary}
\label{se:sum}

If the Higgs boson is not too heavy, an experimental analysis of the
process $\eetth$ at a future high-energy $\Pep\Pem$ linear collider
allows for a determination of the top-quark--Higgs-boson coupling at
the level of $5\%$ or better. At this level of accuracy, the inclusion
of both QCD and electroweak radiative corrections is important.

We have presented the calculation of the ${\cal O}(\alpha)$ and ${\cal
  O}(\alphas)$ radiative corrections to the process $\eetth$ in the
Standard Model in some detail.  Particular attention has been paid to
the treatment of the soft and collinear singularities in the real
photonic (and gluonic) corrections.

Numerical results on the corrections to total cross sections are
discussed, complementing existing work in the literature; in
particular, the interplay of weak fermionic, weak bosonic, and
photonic corrections is investigated.  We find that all these
contributions are typically at the level of $10\%$ and that the
different contributions partially cancel each other.  Moreover, we
have presented numerical results for the electroweak corrections to
differential cross sections, including energy and angular
distributions of the final-state particles.  In the $\GF$-scheme the
size of the corrections typically reaches the $10\%$ level and even
exceeds it near edges of phase space, underlining the importance of a
thorough understanding of the radiative corrections.

\appendix
\section*{Appendix}
\renewcommand{\theequation}{A.\arabic{equation}}

\section{Standard matrix elements}

For the calculation of the loop corrections it is convenient to
separate the matrix elements into scalar invariant coefficient
functions containing the loop integrals and standard matrix elements
(SME) containing all tensorial and spinorial objects and the
dependence on the helicities of the external particles
\cite{Denner:1993kt}.

To introduce a compact notation for the SME, the tensors
\beqar
\Gamma^{\Pep\Pem,\si}_{\{\mu,\mu\nu\rho\}} &=&
\bar v_{\Pep}(p_2)\left\{\ga_\mu,\ga_\mu\ga_\nu\ga_\rho\right\}
\omega_\si u_{\Pem}(p_1),
\nn\\
\Gamma^{\Pt\bar\Pt,\tau}_{\{1,\mu,\mu\nu,\dots\}} &=&
\bar u_\Pt(k_1)\left\{{\bf 1},\ga_\mu,\ga_\mu\ga_\nu,\dots\right\}
\omega_\tau v_{\bar\Pt}(k_2)
\eeqar
are defined with $\omega_\pm=(1\pm\ga_5)/2$ and an obvious notation
for the Dirac spinors $u_{\Pem}(p_1)$, etc. Furthermore, symbols like
$\Gamma_p$ are used as shorthand for the contraction $\Gamma_\mu\,
p^\mu$.  In close analogy to the calculation \cite{Beenakker:2002nc}
of QCD corrections to the related process $q\bar q\to\Pt\bar\Pt\PH$,
we introduce the following set of SME,
\newcommand{\Msme}{\hat\M}
\beqar
\Msme^{\si\tau}_{\{1,2,3,4\}} &=& \Gamma^{\Pep\Pem,\si,\mu} \;
\Gamma^{\Pt\bar\Pt,\tau}_{\{\mu,\mu p_1,\mu p_2,\mu p_1 p_2\}},
\nn\\*
\Msme^{\si\tau}_{\{5,6,7,8\}} &=& \Gamma^{\Pep\Pem,\si,\mu k_1 k_2} \;
\Gamma^{\Pt\bar\Pt,\tau}_{\{\mu,\mu p_1,\mu p_2,\mu p_1 p_2\}},
\nn\\
\Msme^{\si\tau}_{\{9,10,11,12\}} &=& \Gamma^{\Pep\Pem,\si,k_1} \;
\Gamma^{\Pt\bar\Pt,\tau}_{\{1,p_1,p_2,p_1 p_2\}},
\nn\\
\Msme^{\si\tau}_{\{13,14,15,16\}} &=& \Gamma^{\Pep\Pem,\si,k_2} \;
\Gamma^{\Pt\bar\Pt,\tau}_{\{1,p_1,p_2,p_1 p_2\}},
\nn\\
\Msme^{\si\tau}_{\{17,18,19,20\}} &=& \Gamma^{\Pep\Pem,\si,\mu\nu k_1} \;
\Gamma^{\Pt\bar\Pt,\tau}_{\{\mu\nu,\mu\nu p_1,\mu\nu p_2,\mu\nu p_1 p_2\}},
\nn\\
\Msme^{\si\tau}_{\{21,22,23,24\}} &=& \Gamma^{\Pep\Pem,\si,\mu\nu k_2} \;
\Gamma^{\Pt\bar\Pt,\tau}_{\{\mu\nu,\mu\nu p_1,\mu\nu p_2,\mu\nu p_1 p_2\}},
\nn\\*
\Msme^{\si\tau}_{\{25,26,27,28\}} &=& \Gamma^{\Pep\Pem,\si,\mu\nu\rho} \;
\Gamma^{\Pt\bar\Pt,\tau}_{\{\mu\nu\rho,\mu\nu\rho p_1,\mu\nu\rho p_2,
\mu\nu\rho p_1 p_2\}}.
\label{eq:SME}
\eeqar

\begin{sloppypar}
  The four-dimensionality of space--time implies that the SME
  $\Msme^{\si\tau}_{i}$ are not all independent; there are linear
  relations among them.  A simple way to derive the relations with
  real coefficients is provided by the following trick. In four
  dimensions the metric tensor can be decomposed in terms of four
  independent orthonormal four-vectors $n_l$,
\beq
g^{\al\be} = n_0^\al n_0^\be - \sum_{l=1}^3 n_l^\al n_l^\be,
\label{eq:gdecomp2}
\eeq
where $n_k\cdot n_l = g_{kl}$.
Two convenient choices ($j=1,2$) for the vectors $n_l$ are given by
\beqar
n_0^\al &=& \frac{1}{\sqrt{s}}(p_1+p_2)^\al, \qquad
n_1^\al = \frac{1}{\sqrt{s}}(p_1-p_2)^\al,
\nn\\
n_2^\al &=& \sqrt{\frac{s}{\bar t_{1j}\bar t_{2j}-\Mt^2 s}} 
        \left( k_j^\al + \frac{\bar t_{2j}}{s}p_1^\al 
        + \frac{\bar t_{1j}}{s}p_2^\al \right), \quad
\nn\\
n_3^\al &=& \epsilon^{\al\be\ga\de} n_{0,\be}n_{1,\ga}n_{2,\de}=
-\frac{2}{\sqrt{s(\bar t_{1j}\bar t_{2j}-\Mt^2 s)}} \epsilon^{\al\be\ga\de}
        p_{1,\be}p_{2,\ga}k_{j,\de}
\eeqar
with 
\beq
\bar t_{ij} = t_{ij}-\Mt^2, \quad i,j=1,2,
\eeq
and $\epsilon^{\al\be\ga\de}$ ($\epsilon^{0123}=+1$) denoting the
totally antisymmetric tensor. Inserting this decomposition for both
$j=1,2$ in all contractions between different Dirac chains according
to $\Ga^{\Pep\Pem,\si,\al}\Ga^{\Pt\bar\Pt,\tau}_\al =
\Ga^{\Pep\Pem,\si}_{\al}g^{\al\be}\Ga^{\Pt\bar\Pt,\tau}_{\be}$, etc.,
and subsequently using the Dirac equation and the Chisholm identity
\beq
 \ri\epsilon^{\al\be\ga\de}\ga_\de\ga_5=
\ga^\al \ga^\be \ga^\ga - g^{\al\be}\ga^\ga + g^{\al\ga}\ga^\be
- g^{\be\ga}\ga^\al
\label{eq:chisholm}
\eeq
reduces all 112 SME $\Msme^{\si\tau}_{i}$ to 16.  Further relations
with complex coefficients result from the direct application of the
Chisholm identity \refeq{eq:chisholm} to structures like
$\Ga^{\Pep\Pem,\si}_{p_2 k_1 p_1}$ and subsequently using the
decomposition
\beq
g^{\rho\si} = \sum_{i,j=1}^4 \left(Z^{-1}\right)_{ij} 2p_i^\rho p_j^\si,
\qquad Z_{ij} = 2p_i p_j,
\qquad p_3=k_1, \quad p_4=k_2,
\eeq
to separate all contractions between $\eps$ tensors and Dirac chains
via $\Ga^{\Pep\Pem,\si}_{\de} \eps^{\al\be\ga\de} =
\Ga^{\Pep\Pem,\si}_{\rho} g^{\rho\de}
\eps^{\al\be\ga}_{\phantom{\al\be\ga}\de}$.  For
$\Ga^{\Pep\Pem,\pm}_{p_2 k_1 p_1}$, in particular, this leads to
\beq
\Ga^{\Pep\Pem,\si}_{\de} \eps^{p_2 k_1 p_1\de}
= \Ga^{\Pep\Pem,\si}_{\rho} g^{\rho\de} 
\eps^{p_2 k_1 p_1}_{\phantom{p_2 k_1 p_1}\de}
= 2X \left[ \Ga^{\Pep\Pem,\si}_{k_1} \left(Z^{-1}\right)_{43}
+ \Ga^{\Pep\Pem,\si}_{k_2} \left(Z^{-1}\right)_{44} \right]
\eeq
with
\beq
X = \eps^{p_1 p_2 k_1 k_2}
= \eps^{\mu\nu\rho\si} p_{1,\mu} p_{2,\nu} k_{1,\rho} k_{2,\si}.
\eeq
In the inverse matrix $\left(Z^{-1}\right)$, the determinant $\det(Z)$
occurs, which can be identified with $\det(Z)=-16X^2$.
\end{sloppypar}

Altogether, the linear relations reduce the set of SME to 8 SME.  It
is convenient to express all SME in terms of $\Msme^{\si\tau}_1$,
$\Msme^{\pm\pm}_2$, and $\Msme^{\pm\mp}_3$ ($\Msme^{\pm\mp}_2$ and
$\Msme^{\pm\pm}_3$ vanish),
\beqar
\Msme^{{+}{+}}_i &=& 
  a_i  \Msme^{{+}{+}}_1 + b_i  \Msme^{{+}{-}}_1 
+ c_i  \Msme^{{+}{+}}_2 + d_i  \Msme^{{+}{-}}_3,
\nn\\
\Msme^{{-}{-}}_i &=& 
  a_i^*\Msme^{{-}{-}}_1 + b_i^*\Msme^{{-}{+}}_1 
+ c_i^*\Msme^{{-}{-}}_2 + d_i^*\Msme^{{-}{+}}_3,
\nn\\
\Msme^{{+}{-}}_i &=& 
  e_i  \Msme^{{+}{+}}_1 + f_i  \Msme^{{+}{-}}_1 
+ g_i  \Msme^{{+}{+}}_2 + h_i  \Msme^{{+}{-}}_3,
\nn\\
\Msme^{{-}{+}}_i &=& 
  e_i^*\Msme^{{-}{-}}_1 + f_i^*\Msme^{{-}{+}}_1 
+ g_i^*\Msme^{{-}{-}}_2 + h_i^*\Msme^{{-}{+}}_3.
\eeqar
The non-vanishing coefficients $a_i$, etc., for the dependent SME read
\newcommand{\toneone}{\bar t_{11}}
\newcommand{\tonetwo}{\bar t_{12}}
\newcommand{\ttwoone}{\bar t_{21}}
\newcommand{\ttwotwo}{\bar t_{22}}
\newcommand{\sonetwo}{\bar s_{12}}
\arraycolsep 3pt
\def\arraystretch{1.6}
\beq
\lefteqn{
\begin{array}[b]{rclrclrcl}
a_5 &=& -\disp\frac{D^* \Mt^2 - 2 \sonetwo \tonetwo \ttwotwo}{2 E_2}, &
a_6 &=& \disp\frac{\Mt (B^* \toneone - 2 \Mt^2 s \tonetwo)}{2 E_1},
\\[1.5ex]
a_8 &=& -\disp\frac{\ttwotwo(2 \Mt^2 s \toneone - C \tonetwo)}{2 E_2}, &
a_9 &=& \disp\frac{\Mt}{2}, &
a_{10} &=& -\disp\frac{\toneone}{2}, 
\\[1.5ex]
a_{11} &=& \disp\frac{A\ttwotwo}{4 E_2}, &
a_{13} &=& -\disp\frac{A^* \Mt}{4 E_1}, &
a_{14} &=& \disp\frac{A^* \toneone}{4 E_1}, 
\\[1.5ex]
a_{15} &=& -\disp\frac{\ttwotwo}{2}, &
a_{17} &=& 2 \Mt, &
a_{18} &=& -4 \toneone, 
\\[1.5ex]
a_{19} &=& \disp\frac{2 A \ttwotwo}{E_2}, &
a_{21} &=& -\disp\frac{A^* \Mt}{E_1}, &
a_{22} &=& -4 \tonetwo, 
\\
a_{23} &=& -4 \ttwotwo, &
a_{25} &=& 16, 
\\[-2.0em] 
\hphantom{xxx} & \hphantom{x} & \hphantom{xxxxxxxxxxxxxxxxx} & 
\hphantom{xxx} & \hphantom{x} & \hphantom{xxxxxxxxxxxxxxxxx} &
\hphantom{xxx} & \hphantom{x} & \hphantom{xxxxxxxxxxxxxxxxx} 
\earr
}\hspace*{34em}
\eeq
\vspace{.5em}
\beq
\lefteqn{
\begin{array}[b]{rclrclrcl}
b_9 &=& \disp\frac{A \Mt}{4 E_2}, &
b_{12} &=& \disp\frac{A \Mt s}{4 E_2}, &
b_{13} &=& -\disp\frac{\Mt}{2}, 
\\[1.5ex]
b_{16} &=& -\disp\frac{\Mt s}{2}, &
b_{17} &=& \disp\frac{A \Mt}{E_2}, &
b_{20} &=& \disp\frac{A \Mt s}{E_2}, 
\\
b_{21} &=& -2 \Mt, &
b_{24} &=& -2 \Mt s,
\\[-2.0em] 
\hphantom{xxx} & \hphantom{x} & \hphantom{xxxxxxxxxxxxxxxxx} & 
\hphantom{xxx} & \hphantom{x} & \hphantom{xxxxxxxxxxxxxxxxx} &
\hphantom{xxx} & \hphantom{x} & \hphantom{xxxxxxxxxxxxxxxxx} 
\earr
}\hspace{34em}
\eeq
\vspace{.5em}
\beq
\lefteqn{
\begin{array}[b]{rclrclrcl}
c_6 &=& -\disp\frac{A^* \Mt^2}{2 E_1}, &
c_9 &=& \disp\frac{2 \Mt^2 s \ttwoone - B \ttwotwo}{4 E_2 s}, &
c_{10} &=& \disp\frac{\Mt}{2},
\\[1.5ex]
c_{12} &=& -\disp\frac{A \ttwotwo}{4 E_2}, &
c_{13} &=& \disp\frac{C^* \ttwoone - 2 \Mt^2 s \ttwotwo}{4 E_1 s}, &
c_{14} &=& -\disp\frac{A^* \Mt}{4 E_1}, 
\\[1.5ex]
c_{16} &=& \disp\frac{\ttwotwo}{2}, &
c_{17} &=& \disp\frac{2 \Mt^2 s \ttwoone - B \ttwotwo}{E_2 s}, &
c_{20} &=& -\disp\frac{A \ttwotwo}{E_2}, 
\\[1.5ex]
c_{21} &=& \disp\frac{C^* \ttwoone - 2 \Mt^2 s \ttwotwo}{E_1 s}, &
c_{24} &=& 2 \ttwotwo, &
c_{26} &=& 4, 
\\[-2.0em] 
\hphantom{xxx} & \hphantom{x} & \hphantom{xxxxxxxxxxxxxxxxx} & 
\hphantom{xxx} & \hphantom{x} & \hphantom{xxxxxxxxxxxxxxxxx} &
\hphantom{xxx} & \hphantom{x} & \hphantom{xxxxxxxxxxxxxxxxx} 
\earr
}\hspace{34em}
\eeq
\vspace{.5em}
\beq
\lefteqn{
\begin{array}[b]{rclrclrcl}
d_5 &=& \disp\frac{\Mt (2 \Mt^2 s \toneone - C \tonetwo)}{2 E_2 s}, &
d_8 &=& \disp\frac{\Mt (2 \Mt^2 s \toneone - C \tonetwo)}{2 E_2}, &
d_{11} &=& -\disp\frac{A \Mt}{4 E_2},
\\[1.5ex]
d_{15} &=& \disp\frac{\Mt}{2}, &
d_{19} &=& -\disp\frac{2 A \Mt}{E_2}, &
d_{23} &=& 4 \Mt, 
\\[-2.0em] 
\hphantom{xxx} & \hphantom{x} & \hphantom{xxxxxxxxxxxxxxxxx} & 
\hphantom{xxx} & \hphantom{x} & \hphantom{xxxxxxxxxxxxxxxxx} &
\hphantom{xxx} & \hphantom{x} & \hphantom{xxxxxxxxxxxxxxxxx} 
\earr
}\hspace{34em}
\eeq
\vspace{.5em}
\beq
\lefteqn{
\begin{array}[b]{rclrclrcl}
e_6 &=& -\disp\frac{\Mt (2 \Mt^2 s \toneone - C \tonetwo)}{2 E_2}, &
e_9 &=& \disp\frac{A \Mt}{4 E_2}, &
e_{13} &=& -\disp\frac{\Mt}{2}, 
\\[1.5ex]
e_{17} &=& \disp\frac{2 A \Mt}{E_2}, &
e_{21} &=& -4 \Mt,
\\[-2.0em] 
\hphantom{xxx} & \hphantom{x} & \hphantom{xxxxxxxxxxxxxxxxx} & 
\hphantom{xxx} & \hphantom{x} & \hphantom{xxxxxxxxxxxxxxxxx} &
\hphantom{xxx} & \hphantom{x} & \hphantom{xxxxxxxxxxxxxxxxx} 
\earr
}\hspace{34em}
\eeq
\vspace{.5em}
\beq
\lefteqn{
\begin{array}[b]{rclrclrcl}
f_4 &=& s, &
f_5 &=& -\disp\frac{A^* \Mt^2}{2 E_1}, &
f_8 &=& -\disp\frac{A^* \Mt^2 s}{2 E_1}, 
\\[1.5ex]
f_9 &=& \disp\frac{\Mt}{2}, &
f_{10} &=& \disp\frac{A \tonetwo}{4 E_2}, &
f_{11} &=& -\disp\frac{\ttwoone}{2}, 
\\[1.5ex]
f_{12} &=& \disp\frac{\Mt s}{2}, &
f_{13} &=& -\disp\frac{A^* \Mt}{4 E_1}, &
f_{14} &=& -\disp\frac{\tonetwo}{2}, 
\\[1.5ex]
f_{15} &=& \disp\frac{A^* \ttwoone}{4 E_1}, &
f_{16} &=& -\disp\frac{A^* \Mt s}{4 E_1}, &
f_{18} &=& \disp\frac{A \tonetwo}{E_2},
\\[1.5ex]
f_{19} &=& -2 \ttwoone, &
f_{22} &=& -2 \tonetwo, &
f_{23} &=& \disp\frac{A^* \ttwoone}{E_1},
\\
f_{25} &=& 4, &
f_{28} &=& 4s,
\\[-2.0em] 
\hphantom{xxx} & \hphantom{x} & \hphantom{xxxxxxxxxxxxxxxxx} & 
\hphantom{xxx} & \hphantom{x} & \hphantom{xxxxxxxxxxxxxxxxx} &
\hphantom{xxx} & \hphantom{x} & \hphantom{xxxxxxxxxxxxxxxxx} 
\earr
}\hspace{34em}
\eeq
\vspace{.5em}
\beq
\lefteqn{
\begin{array}[b]{rclrclrcl}
g_{10} &=& -\disp\frac{A \Mt}{4 E_2}, &
g_{14} &=& \disp\frac{\Mt}{2}, &
g_{18} &=& -\disp\frac{A \Mt}{E_2}, 
\\
g_{22} &=& 2 \Mt, 
\\[-2.0em] 
\hphantom{xxx} & \hphantom{x} & \hphantom{xxxxxxxxxxxxxxxxx} & 
\hphantom{xxx} & \hphantom{x} & \hphantom{xxxxxxxxxxxxxxxxx} &
\hphantom{xxx} & \hphantom{x} & \hphantom{xxxxxxxxxxxxxxxxx} 
\earr
}\hspace{34em}
\eeq
\vspace{.5em}
\beq
\lefteqn{
\begin{array}[b]{rclrclrcl}
h_5 &=& \disp\frac{\Mt (B^* \toneone - 2 \Mt^2 s \tonetwo)}{2 E_1 s}, &
h_6 &=& \disp\frac{\tonetwo (2 \Mt^2 s \toneone - C \tonetwo)}{2 E_2 s}, &
h_7 &=& \disp\frac{B^*}{2s}, 
\\[1.5ex]
h_8 &=& \disp\frac{\Mt (B^* \toneone - 2 \Mt^2 s \tonetwo)}{2 E_1}, &
h_9 &=& \disp\frac{2 \Mt^2 s \toneone - C \tonetwo}{4 E_2 s}, &
h_{11} &=& \disp\frac{\Mt}{2}, 
\\[1.5ex]
h_{12} &=& -\disp\frac{\toneone}{2}, &
h_{13} &=& \disp\frac{B^* \toneone - 2 \Mt^2 s \tonetwo}{4 E_1 s}, &
h_{15} &=& -\disp\frac{A^* \Mt}{4 E_1}, 
\\[1.5ex]
h_{16} &=& \disp\frac{A^* \toneone}{4 E_1}, &
h_{17} &=& \disp\frac{2 (2 \Mt^2 s \toneone - C \tonetwo)}{E_2 s}, &
h_{19} &=& 2 \Mt, 
\\[1.5ex]
h_{20} &=& -4 \toneone, &
h_{23} &=& -\disp\frac{A^* \Mt}{E_1}, &
h_{24} &=& -4 \tonetwo, 
\\
h_{27} &=& 16, 
\\[-2.0em] 
\hphantom{xxx} & \hphantom{x} & \hphantom{xxxxxxxxxxxxxxxxx} & 
\hphantom{xxx} & \hphantom{x} & \hphantom{xxxxxxxxxxxxxxxxx} &
\hphantom{xxx} & \hphantom{x} & \hphantom{xxxxxxxxxxxxxxxxx} 
\earr
}\hspace{34em}
\eeq
where the following shorthands are used,
\beqar
A  &=&  s \sonetwo - \tonetwo \ttwoone - \toneone \ttwotwo + 4 \ri X,
\qquad
B \;=\; s \sonetwo + \tonetwo \ttwoone - \toneone \ttwotwo + 4 \ri X,
\nn\\
C  &=&  s \sonetwo - \tonetwo \ttwoone + \toneone \ttwotwo + 4 \ri X,
\qquad
D \;=\; s \sonetwo + \tonetwo \ttwoone + \toneone \ttwotwo + 4 \ri X,
\nn\\
E_i &=& -\Mt^2  s + \bar t_{1i} \bar t_{2i}, \qquad i=1,2,
\eeqar
and
\beq
\bar s_{12} = s_{12}-2\Mt^2. \qquad
\eeq

Finally, we give the explicit expressions of the independent SME
$\Msme^{\si\tau}_1$, $\Msme^{\pm\pm}_2$, and $\Msme^{\pm\mp}_3$   
in terms of Weyl--van der Waerden spinor products as introduced in \refse{se:convs},
\beq
\begin{array}[b]{rclrcl}
\Msme^{++}_1 &=& 2\Cptxi\poXI, & \qquad 
\Msme^{--}_1 &=& 2\CpoETA\pteta, \\ 
\Msme^{+-}_1 &=& 2\CptETA\poeta, &  
\Msme^{-+}_1 &=& 2\Cpoxi\ptXI, \\
\Msme^{++}_2 &=& 2\Cpopt\poXI\poeta, & \qquad
\Msme^{--}_2 &=& -2\Cpoxi\CpoETA\popt, \\    
\Msme^{+-}_3 &=& 2\Cptxi\CptETA\popt, & 
\Msme^{-+}_3 &=& -2\Cpopt\ptXI\pteta, 
\end{array}
\eeq
where the insertions for the spinors $\xi^{(\prime)}$ and
$\eta^{(\prime)}$ for the polarizations $\tau_{1,2}$ are specified in
Eq.~\refeq{eq:etaxi}.

\renewcommand{\theequation}{B.\arabic{equation}}

\section{Lowest-order and one-loop amplitudes in terms of
standard matrix elements}

Both lowest-order and one-loop amplitudes can be written as linear
combination of the SME defined in Eq.~\refeq{eq:SME}.
The contributions to the lowest-order amplitude $\M_0$, as defined
in Eq.~\refeq{eq:MEttH}, take the simple form
\beqar\label{eq:MEttHii}
\M_0^{\ZH,\si} &=&  \sum_{\tau=\pm}
e^3 \frac{\gZes\, g_{\PZ\PZ\PH}\,\gZtt}{(s-\MZ^2)(s_{12}-\MZ^2)} 
\Msme^{\si\tau}_1,
\nn
\\[.5em]
\M_0^{\XH,\si} &=&  \sum_{\tau=\pm}
2{e^3} \frac{\gZes\, g_{\PZ\chi\PH}\,g_{\chi\Pt}^\tau}{(s-\MZ^2)(s_{12}-\MZ^2)} 
(\Msme^{\si\tau}_9+\Msme^{\si\tau}_{13}),
\nn
\\[.5em]
\M_0^{\VtH,\si} &=&  \sum_{\tau=\pm}
e^3 \frac{\gVes\,\gVtt\, g_{\PH\Pt}}{(s-\MV^2)(s_{13}-\Mt^2)} 
\nn\\
&& {} \times
\left[\Msme^{\si\tau}_2+\Msme^{\si\tau}_3+2\Msme^{\si\tau}_{13}
-\Mt(\Msme^{\si\tau}_1-\Msme^{\si,-\tau}_1)\right],
\nn\\[.5em]
\M_0^{\VtbarH,\si} &=&  \sum_{\tau=\pm}
{e^3}\frac{\gVes\,\gVtt\, g_{\PH\Pt}}{(s-\MV^2)(s_{23}-\Mt^2)} 
\nn\\
&& {} \times
\left[\Msme^{\si,-\tau}_2+\Msme^{\si,-\tau}_3-2\Msme^{\si,-\tau}_9
-\Mt(\Msme^{\si\tau}_1-\Msme^{\si,-\tau}_1)\right].
\eeqar
The one-loop amplitude $\M^\si_1$ 
is much too lengthy to be written down; it is of the form
\beq\label{eq:mefromsme}
\M_1^\si = \sum_{i=1}^{28} \sum_{\tau=\pm} \,
F^{\si\tau}_i \Msme^{\si\tau}_i,
\eeq
where the coefficient functions $F^{\si\tau}_i$ depend on couplings
and scalar products of external momenta and contain all loop integrals.
Note that here 
$\tau$ is related to the chirality projectors 
$\om_\tau$, whereas in \refse{se:convs} $\tau_1,\tau_2$ denote the
helicities of the external fermions. 

\section*{Acknowledgements}

This work was supported in part by the Swiss Bundesamt f\"ur Bildung
und Wissenschaft and by the European Union under contract HPRN-CT-2000-00149.

\end{document}